\documentclass[10pt,letterpaper]{article}
\usepackage[margin=1in,nohead,centering]{geometry}
\usepackage{url}
\usepackage{xspace,graphics,graphicx}
\usepackage{amsmath,amssymb}

\newcommand{\aseq}{{\bf a}\xspace}

\newcommand{\op}{\,\mbox{\bf\texttt{(}}\,}
\newcommand{\cp}{\,\mbox{\bf\texttt{)}}\,}

\title{RNA thermodynamic structural entropy}
\author{J.A. Garcia-Martin,
P. Clote\thanks{Corresponding author.}
}
\date{Biology, Boston College\\ 
Chestnut Hill, MA 02467 \\
{\tt garcaama@bc.edu}
{\tt clote@bc.edu}
}
\begin{document}
\maketitle

\begin{abstract}
Conformational entropy for atomic-level, three dimensional biomolecules is
known experimentally to play an important role in protein-ligand 
discrimination, yet reliable computation of entropy remains a difficult
problem.
Here we describe the first two accurate and efficient algorithms to compute 
the conformational entropy for RNA secondary structures, with respect 
to the Turner energy model, where free energy parameters are determined 
from UV aborption experiments. An algorithm to compute the derivational 
entropy 
for RNA secondary structures had previously been introduced, using
stochastic context free grammars (SCFGs).  However, the numerical value 
of derivational entropy depends heavily on the chosen context free grammar 
and on the training set
used to estimate rule probabilities.  Using data from the Rfam database,
we determine that both of our thermodynamic methods, which agree in 
numerical value, are substantially faster than the SCFG method.
Thermodynamic structural entropy is much smaller than 
derivational entropy, and the correlation between 
length-normalized thermodynamic entropy and derivational entropy 
is moderately weak to poor.  In applications, we
plot the structural entropy as a function of temperature for known
thermoswitches, such as the 
repression of heat shock gene expression (ROSE) element,
we determine that the correlation between hammerhead ribozyme cleavage
activity and total free energy is improved by including an additional
free energy term arising from conformational entropy, and we
plot the structural entropy of windows of the HIV-1 genome.

Our software {\tt RNAentropy} can compute
structural entropy for any user-specified temperature, and supports both
the Turner'99 and Turner'04 energy parameters. It follows that
{\tt RNAentropy} is state-of-the-art software to compute RNA secondary
structure conformational entropy.  The software is available at
\url{http://bioinformatics.bc.edu/clotelab/RNAentropy}.
\end{abstract}

\section*{Introduction}
\label{section:intro}
Conformational (or configurational) entropy is defined by
\begin{eqnarray}
\label{eqn:conformationalEntropy}
S = -k_B \sum_s p(s) \ln p(s)
\end{eqnarray}
where $k_B$ denotes the Boltzmann constant, and the sum is taken over
all structures.  As shown experimentally to be the case for
calmodulin \cite{Marlow.ncb10}, conformational 
entropy plays an important role for the discrimination
observed in protein-ligand binding. Since conformational entropy is
well-known to be difficult to measure, this recent experimental advance
involves using NMR relaxation as a 
proxy for entropy, a technique reviewed in \cite{Wand.cosb13}.

It is currently not possible to reliably compute the conformational entropy for
3-dimensional molecular structures \cite{Wand.cosb13}; nevertheless,
various methods have been developed, employing approaches from
molecular, harmonic, and quasiharmonic dynamics
\cite{Karplus.bj87,Harpole.jpcb11}.
It appears likely that such computational methods will continue to improve,
especially with the availability now of experimentally determined values
by using NMR relaxation \cite{Wand.cosb13}.

In contrast to the complex situation for 3-dimensional molecular structures,
we show here that it is possible to accurately and efficiently 
compute the exact value of conformational
entropy for RNA secondary structures, with respect to the Turner energy 
model \cite{Turner.nar10},
whose free energy parameters are experimentally determined from
UV absorption experiments \cite{tinoco:reviewInBook}. 
Our resulting
algorithm, {\tt RNAentropy}, runs in cubic time with quadratic memory
requirements, thus answering a question raised by M. Zuker (personal
communication, 2009).

The {\em nearest neighbor} or {\em Turner} energy model is a coarse-grained 
RNA secondary structure model that includes free energy parameters for 
base stacking and various loops (hairpins, bulges, internal loops, 
multiloops) \cite{Turner.nar10}. The exact definition of these loops
can be found in the description of Zuker's algorithm \cite{zukerStiegler}
which computes the minimum free energy (MFE) secondary structure with respect
to the Turner energy model.
%Following pioneering work of Tinoco et al. \cite{Tinoco.n71}, 
%in a series of publications \cite{Freier.pnas86}, D.H. Turner and co-workers
As explained in \cite{tinoco:reviewInBook}, values for base stacking
enthalpy and entropy can be determined by plotting the experimentally measured
UV absorption values of various double-stranded RNA oligonucleotide sequences
at 280 nm (also 260 nm) as a function of RNA concentration. 
By least-squares fitting of
the data, free energy parameters for base stacking, hairpins, bulges, etc.
can be determined. Free energy and enthalpy parameters for an earlier model
(Turner 1999) and a more recent model (Turner 2004)
are described at the Nearest Neighbor Database (NNDB) \cite{Turner.nar10}.
For instance, the base stacking free energy for 
$\begin{array}{ll} 
\mbox{$5'$-{\tt GC}-$3'$}\\ 
\mbox{$3'$-{\tt CG}-$5'$}\\ 
\end{array}$ 
is $-3.4$ kcal/mol in the Turner 2004 parameter set.
MFOLD \cite{zuker:mfoldWebserver}, UNAFOLD \cite{Markham.mmb08}
and the Vienna RNA Package \cite{Lorenz.amb11} are software packages
that implement the Zuker dynamic programming algorithm \cite{zukerStiegler}
to compute the MFE structure as well as the McCaskill algorithm 
\cite{mcCaskill} to compute the partition function over all secondary
structures. Applications of such software are far-reaching, ranging from
the prediction of microRNA target sites \cite{Hammell.nm08} to 
the design of synthetic RNA \cite{Choi.nb10,Dotu.nar15}.

Throughout this paper, for a given RNA sequence $\aseq=a_1,\ldots,a_n$,
{\em structural entropy}, denoted by $H(\aseq)$, is defined to be (Shannon)
entropy
\begin{eqnarray}
\label{eqn:shannonEntropy}
H(\aseq) = -\sum_s p(s) \ln p(s)
\end{eqnarray}
where the sum is taken over all secondary structures $s$ of $\aseq$,
$p(s)$ denotes the Boltzmann probability $\exp(-E(\aseq,s)/RT)/Z(\aseq)$,
$R$ denotes the universal gas constant (Boltzmann constant times Avagadro's
number), $E(\aseq,s)$ is the free energy of the secondary structure $s$ of
$\aseq$ with respect to the Turner energy model \cite{Turner.nar10},
and $Z(\aseq)$ denotes the partition function, defined as the sum of all
Boltzmann factors $\exp(-E(\aseq,s)/RT)$ over all secondary structures $s$.
When the RNA sequence $\aseq$ is clear from the context, we generally write
$E(s)$, $H$ and $Z$, rather than $E(\aseq,s)$, $H(\aseq)$ and $Z(\aseq)$.
It follows that the conformational entropy is equal to the Boltzmann constant
times the structural entropy: $S = k_B H$.

Before presenting our results and methods, we first survey several distinct
notions of entropy that have appeared in 
the literature of RNA secondary structures -- each quite different from
the notion of thermodynamic structural entropy described in this paper.

\subsection*{Pointwise entropy in multiple alignments}

Shannon entropy is used to quantify the variability of positions in 
a multiple sequence alignment. This application is particularly widespread
due to the ubiquitous use of sequence logos \cite{Schneider.nar90,Crooks.gr04} 
to present motifs in proteins, DNA and RNA.
Letting $\mathbb{A}$ denote the 4-letter alphabet $\{A,C,G,U\}$, the pointwise 
entropy $H_1(k)$ at position $k$ in the alignment is defined by
$H_1(k) = -\sum_{a \in \mathbb{A}} p_a \ln p_a$, where $p_a$ is the proportion
of nucleotide $a$ at position $k$.  Entropy values range from 
$0$ to $\log 4$, where
high entropy entails uncertainty or disagreement of the nucleotides at
position $k$. Average pointwise sequence entropy is often expressed in
bits, where logarithm base 2 is used instead of the natural logarithm.
The concept of sequence logo has many generalizations; indeed, logos for
DNA major groove binding are described in \cite{Crooks.gr04},
logos for tertiary structure alignment of proteins are described in
\cite{Bindewald.nar06},
logos for RNA alignments including mutual information on base pair
covariation are described in \cite{Gorodkin.cab97},
and logos with secondary structure context
of RNAs that bind to specific riboproteins are described in
\cite{Kazan.pcb10,Kazan.nar13}.

\subsection*{Positional entropy}

For a given RNA sequence $\aseq = a_1,\ldots,a_n$, and for $1 \leq i<j \leq n$,
define the base pairing probability $p_{i,j}$ to be the sum of
Boltzmann factors of all secondary structures that contain base pair 
$(i,j)$, divided by the partition function, i.e.
\begin{eqnarray}
\label{def:pij}
p_{i,j} = \displaystyle\sum_{\{ s: (i,j) \in s\}} p(s) =
\frac{\sum_{\{ s: (i,j) \in s\}} \exp(-E(s)/RT) }{Z}
\end{eqnarray}
Here $p(s)$ is the Boltzmann probability of structure $s$ of $\aseq$,
$E(s)$ is the Turner free energy of secondary structure $s$ \cite{Turner.nar10},
$R \approx 0.001987$ kcal/mol.K is the universal gas constant, $T$ is
absolute temperature, and the {\em partition function}
$Z = \sum_{s} \exp(-E(s)/RT)$, where the sum is taken over all
secondary structures $s$ of $\aseq$. Base pairing probabilities
can be computed in cubic time by McCaskill's algorithm \cite{mcCaskill},
as implemented in various software, including the Vienna RNA Package
{\tt RNAfold -p} \cite{Lorenz.amb11}. 

Define the positional base pairing probability distribution at
fixed position $1 \leq i \leq n$ by
\begin{eqnarray}
\label{def:pijstar}
p^*_{i,j} = \left\{ \begin{array}{ll}
p_{i,j} &\mbox{if $i<j$}\\
p_{j,i} &\mbox{if $i>j$}\\
1-\displaystyle\sum_{j \ne i} p^*_{i,j} &\mbox{if $i=j$}
\end{array} \right.
\end{eqnarray}
For each fixed value of $i$, $p^*_{i,j}$ is a probability distribution,
where $j$ ranges over $1,\ldots,n$, the positional structural
entropy $H_2(i)$ at position $i$ is defined by
\begin{eqnarray}
\label{def:positionalStrEntropy}
H_2(i) =  -\sum_{j=1}^n p^*_{i,j} \ln p^*_{i,j}.
\end{eqnarray}
Low values of positional entropy at position $i$ indicate that
there is a strong agreement among low energy structures in the Boltzmann
ensemble that either $i$ is unpaired, or that $i$ is paired with the same
position $j$. The {\em average positional entropy} 
$\langle H_2 \rangle$ is the average $\sum_{i=1}^n \frac{H_2(i)}{n}$ taken
over all positions of the sequence. Structural positional entropy was
first defined by Huynen et al. \cite{Huynen.jmb97}, who used the term
$S$-value for average positional entropy, and 
showed that RNA nucleotide positions having low entropy correspond to
positions where the minimum free energy (MFE) structure tends to agree with that
determined by comparative sequence analysis. In \cite{mathews:PairProb},
Mathews made a similar analysis, where in place of $S$-value, a 
normalized pseudo-entropy value was used, defined by
$- \sum_{1 \leq i<j \leq n} p_{i,j} \ln p_{i,j}/n$.
Positional entropy of RNA secondary structures can be presented by
color-coding each nucleotide, where the color of the $k$th nucleotide
reflects the positional entropy $H_2(k)$ as defined in 
equation~(\ref{def:positionalStrEntropy}). The Rfam 12.0 database 
\cite{Nawrocki.nar14} uses such color-coded secondary structures,
since the base-pairing of positions having low entropy is
likely to be correct \cite{Huynen.jmb97,mathews:PairProb}.

\subsection*{Derivational entropy using stochastic context free grammars}
\label{section:SCFG}

Manzourolajdad et al.  \cite{Manzourolajdad.jtb13},
Sukosd et al.  \cite{Sukosd.bb13} and Anderson et al. \cite{Anderson.bb13} 
describe the computation of structural entropy for stochastic
context free grammars (SCFGs), defined by $-\sum_s p(s) \ln p(s)$,
where the sum is taken over all secondary structures $s$ of a given
RNA sequence, and $p(s)$ is the probability of deriving the structure $s$
in a particular grammar $G$, defined as follows.
Suppose that $S=S_0$ is the starting nonterminal for the
grammar $G$, $s=S_m$ is the secondary structure $s$ consisting only of
terminal symbols belonging to the alphabet $\{\op, \cp,\bullet\}$, and
that $S_1,\ldots,S_{m-1}$ are expressions consisting of a mix of nonterminal
and terminal symbols. If
$S_0 \rightarrow_G S_1 \rightarrow_G S_2 \rightarrow_G \cdots \rightarrow_G 
S_m$ is a leftmost derivation using production rules from grammar $G$  and
for each $i=0,\ldots,m-1$, we let $p_i$ denote the probability of applying the 
rule $S_i \rightarrow S_{i+1}$, then $p(s)$ is defined to be the
product $\prod_{i=0}^{m-1} p_i$. It should be noted that the derivational
probability $p(s)$ heavily depends on the choice of grammar $G$ as well
as on the rule application probabilities $p_i$, obtained by 
applying expectation maximization to a chosen
training set of secondary structures.

Anderson et al. \cite{Anderson.bb13} are motivated to compute
derivational entropy of a multiple alignment of RNAs, 
in order to provide a numerical quantification for the
quality of the alignment -- specifically, their paper
shows that accurate alignment quality corresponds to low derivational entropy. 
In \cite{Sukosd.bb11},
Sukosd et al. describe the software {\tt PPfold}, a multithreaded version 
of the {\tt Pfold} RNA secondary structure prediction algorithm.
Subsequently, Sukosd et al. \cite{Sukosd.bb13} describe how to compute
the derivational entropy for the grammar used in the {\tt PFold} algorithm
(grammar G6 as defined in \cite{Dowell.bb04}), and show that derivational
entropy is correlated with the accuracy of {\tt PPfold} structure
predictions, as measured by F-scores.  In contrast,
Manzourolajdad et al.  \cite{Manzourolajdad.jtb13} computed the
derivational entropy of various families of noncoding RNAs, using
the trained stochastic context free grammars G4,G5,G6 \cite{Dowell.bb04},
which they denote respectively as RUN (G4), IVO (G5) and BJK (G6).
The Linux executable and trained models
can be downloaded from \url{http://rna-informatics.uga.edu/malmberg/}
for three RNA stochastic context free grammars, each with three trained
models using the training sets `Rfam5', `Mixed80', and `Benchmark' --
see \cite{Manzourolajdad.jtb13} for description.

The plan of the remainder of this paper is as follows. 
In Section~\ref{section:methods}, we provide a description of
our novel entropy algorithms, beginning with an overview in
Section~\ref{section:statisticalMechanics}, where
we derive a relation between structural entropy $H$ and expected energy
$\langle E \rangle$. This relation allows us to provide a crude estimate of $H$
by sampling. Expected energy can be computed from the derivative of the 
logarithm
of the partition function with respect to temperature; a finite difference
computation then yields our first algorithm to compute structural entropy,
while a dynamic programming approach for the expected energy yields our second
algorithm. In Section~\ref{section:results}, we compare our structural entropy
software, {\tt RNAentropy}, with software for SCFG
derivational entropy, and then use {\tt RNAentropy} in several applications.
Section~\ref{section:comparison} benchmarks the time
required to compute structural entropy using our two algorithms, versus
the time required to compute derivational entropy using
the program of \cite{Manzourolajdad.jtb13}. Numerical values for
structural and derivational entropies are compared, along with their 
distributions.  In addition
to a comparison of run times and entropy values,
we compute the correlation of structural entropy, derivational entropy,
and a variety of measures,
such as ensemble defect \cite{Dirks.nar04}, positional
entropy \cite{Huynen.jmb97}, structural diversity \cite{morganHiggsBarrier},
etc. Such measures have recently been used in the design of experimentally
validated RNA molecules \cite{Zadeh.jcc11,Dotu.nar15}.
Motivated by the fact that calmodulin-ligand binding has been shown to
depend on conformational entropy \cite{Marlow.ncb10}, 
in Section~\ref{section:hammerheadActivity}, we  show an improvement in
the correlation between hammerhead ribozyme cleavage activity and 
total change of energy \cite{Shao.bb07}, 
if conformational entropy is also taken into account.
In Section~\ref{section:hiv}, we compute the entropy of genomic portions of
the HIV-1 genome and compare entropy Z-scores with known HIV-1 noncoding
elements. Finally, in Section~\ref{section:discussion},
we describe differences between the methods and discuss the
numerical discrepancy between thermodynamic structural
entropy values and SCFG derivational entropy values.  For more background 
on RNA, an excellent, though somewhat outdated, review of computational and 
physical aspects of RNA is given by Higgs \cite{Higgs.qrb00}.

% You may title this section "Methods" or "Models". 
% "Models" is not a valid title for PLoS ONE authors. However, PLoS ONE
% authors may use "Analysis" 
\section*{Methods}
\label{section:methods}

In this section, we describe the two novel algorithms to compute
RNA thermodynamic structural entropy using the Turner energy model
\cite{Turner.nar10}. Section \ref{section:statisticalMechanics}
describes the relation between entropy and expected energy, and provides
two variants of a simple sampling method to approximate the value of
structural entropy. The approximation does not yield accurate entropy
values, so two accurate methods are described: (1) formal temperature
derivative (FTD) method, (2) dynamic programming (DP) method. An overview
of both algorithms is provided in this section.  Full details of each
algorithm are then provided in 
Sections~\ref{section:statisticalPhysics} and \ref{section:entropyDP}.

\subsection*{Statistical mechanics}
\label{section:statisticalMechanics}

Shannon entropy for the Boltzmann ensemble of secondary structures of a
given RNA sequence $\aseq = a_1,\ldots,a_n$ is defined by
\begin{eqnarray}
\label{eqn:eqnShannonEntropyDef}
H(\aseq) &=& - \sum_s p(s) \ln p(s) =
 - \sum_s \frac{\exp(-E(s)/RT)}{Z}
 \ln \left( \frac{\exp(-E(s)/RT)}{Z} \right)  \nonumber \\
 &=& - \sum_s \frac{\exp(-E(s)/RT)}{Z} \cdot \left[
-\frac{E(s)}{RT} - \ln Z \right] \nonumber \\
 &=& \frac{1}{RT} \sum_s  p(s) E(s) + \frac{\ln Z}{Z} \cdot
\sum_{s} \exp(-E(s)/RT) \nonumber \\
&=&  \frac{\langle E \rangle}{RT}  + \ln Z = \frac{\langle E \rangle - G}{RT}
\end{eqnarray}
where $G$ denotes the ensemble free energy $-RT \ln Z$.
It follows that if the energy $E(s)$ of every structure $s$ is zero, or
if the temperature $T$ is infinite, then entropy is equal to
the logarithm of the number of structures.  Note as well that in the
Nussinov energy model \cite{nussinovJacobson}, where each base pair
has an energy of $-1$, it follows that
the expected energy is equal to $-1$ times the expected number of base
pairs, i.e.  $\langle E \rangle = - \sum_{i<j} p_{i,j}$, where
$p_{i,j}$ is the probability of base pair $(i,j)$ in the Nussinov model.

By sampling RNA structures with the {\tt RNAsubopt} program from
Vienna RNA Package \cite{Lorenz.amb11}, we can approximate the value of 
expected energy, and hence obtain an approximation of the thermodynamic 
entropy by using equation~(\ref{eqn:eqnShannonEntropyDef}). 
This can be done in two distinct manners.

In the first approach, a user-specified number {\tt N} of 
low energy structures from the thermodynamic
ensemble can be sampled by using the algorithm of Ding and
Lawrence \cite{Ding.nar03}, as implemented in {\tt RNAsubopt -p N}. A sampling
approximation for the expected energy is then defined to be the
arithmetic average of the free energy of the {\tt N} sampled structures.
In the second approach, all structures can be generated, whose free energy 
lies within a user-specified range {\tt E} of the minimum free energy,
by using the algorithm of Wuchty \cite{wuchtyFontanaHofackerSchuster},
as implemented in
{\tt RNAsubopt -e E}. Let $Z_0$ be an approximation of the partition
function, defined by summing the Boltzmann factors $\exp(-E(s)/RT)$
for all generated structures. Define the (approximate) Boltzmann probability
of a generated structure $s$ to be $p(s)=\exp(-E(s)/RT)/Z_0$.
An approximation for the expected energy is in this case taken to be 
$\sum_{s} p(s) \cdot E(s)$, where the sum is taken over all structures
$s$, whose free energy is within {\tt E}  kcal/mol of the minimum free energy.
In either case, the resulting entropy approximation is not particularly
good. For instance, the thermodynamic entropy of the 78 nt arginyl-tRNA
from {\em Aeropyrum pernix} (accession code
tdbR00000589 in the {\em Transfer RNA database} tRNAdb \cite{Juhling.nar09})
is 5.44, as computed by the algorithm {\tt RNAentropy} described in this
paper, while the entropy approximation by the first sampling approach 
with $N=10,000$ is
4.71 and that of the second sampling approach with $E=10$ is 4.68. 
%Our sampling code is available at the web server for {\tt RNAentropy}. 
Since the estimate from each sampling approach has greater than 13\%
relative error, sampling cannot be used to provide accurate entropy
values. For that reason, we now briefly describe two novel, cubic time
algorithms to compute the exact value of structural entropy -- details of
the algorithms are further described in 
Sections~\ref{section:statisticalPhysics} and \ref{section:entropyDP}.

\subsection*{Algorithm 1: Formal temperature derivative (FTD)}

It is well-known from statistical physics that the average energy 
$\langle E \rangle$ of $N$ independent and distinguishable particles is
given by the following formula (cf equation (10.36) of \cite{dillBromberg}):
\begin{eqnarray}
\label{eqn:expEnergyDill}
\langle E \rangle &=& RT^2 \cdot \frac{\partial}{\partial T} \ln Z(T).
\end{eqnarray}
This equation does not hold in the case of RNA secondary structures with
the Turner energy model; however, 
equation~(\ref{eqn:expEnergyDill}) is close to being correct.
The idea of Algorithm 1 is to use finite differences
$\frac{\ln Z(T+\Delta T) - \ln Z(T)}{\Delta T}$ to approximate the
derivative $\frac{\partial}{\partial T} \ln Z(T)$, thus obtaining
the expected energy $\langle E \rangle$, from which we obtain the
structural entropy by applying equation~(\ref{eqn:eqnShannonEntropyDef}). 
As shown later, certain technically subtle issues arise in this approach;
in particular, the derivative
$\frac{\partial}{\partial T} \ln Z(T)$ must be taken with respect to
the {\em formal temperature}, which represents only those occurrences
of the temperature variable within the expression $RT$. Formal temperature
is distinct from {\em table temperature}, which latter designates all
occurrences of the temperature variable in the Turner energy parameters.
This will be fully explained in Section~\ref{section:statisticalPhysics}.
For this
reason, Algorithm 1 is named {\tt FTD}, for formal temperature derivative.

\subsection*{Algorithm 2: Dynamic Programming (DP)}

Recall that the partition function for a given RNA sequence $\aseq$ 
is defined by
$Z = \sum_s \exp(-E(s)/RT)$, where the sum is taken
over all secondary structures of ${\bf a}$. Letting 
$BF(s) = \exp(-E(s)/RT)$ denote the Boltzmann factor of $s$, it follows
that the Boltzmann probality of secondary structure $s$ satisfies
$p(s)=BF(s)/Z$, and hence
\[
\langle E \rangle = \sum_s p(s) \cdot E(s) 
= \sum_{s} \frac{BF(s) \cdot E(s)}{Z} = \frac{Q}{Z}
\]
where $Q = \sum_s BF(s) \cdot E(s)$.
The partition function $Z$ can be computed by McCaskill's algorithm
\cite{mcCaskill}, while in
Sections~\ref{section:entropyDP}, we describe a dynamic programming
algorithm to compute $Q(a)$. 
Since this method uses dynamic programming, Algorithm 2 is named {\tt DP}.

Both FTD and DP support the Turner'99 and Turner'04 energy models
\cite{Turner.nar10}, and all references to FTD and DP mean
FTD'04 and DP'04, unless otherwise stated (there are small numerical
differences in the entropy, depending on the choice of Turner parameters).
Moreover, both algorithms allow the user to specify an arbitrary temperature
$T$ for the computation of structural entropy. This latter feature could
prove useful in the investigation of thermoswitches, also called RNA
thermometers, discussed later. 
The software {\tt RNAentropy} implements both algorithms, and is available
at \url{http://bioinformatics.bc.edu/clotelab/RNAentropy}.

\subsection*{Entropy by statistical physics}
\label{section:statisticalPhysics}

Here we show that for the Turner energy model of RNA secondary structures,
expected energy satisfies
\begin{eqnarray}
\label{eqn:expEnergyApproxEqualDerivative}
\langle E \rangle \approx RT^2 \cdot \frac{\partial}{\partial T} \ln Z(T)
\end{eqnarray}
although equality does not strictly hold. Indeed,
\begin{eqnarray}
\label{eqn:expEnergyStatPhysics}
RT^2 \cdot \frac{\partial}{\partial T} \ln Z(T) &=& \frac{RT^2}{Z(T)}
\cdot \frac{\partial}{\partial T} Z(T) 
= \frac{RT^2}{Z(T)} \sum_{s \in \mathbb{SS}(\aseq)}
\frac{\partial}{\partial T} \exp(-E(s)/RT) \nonumber \\
&=& \frac{RT^2}{Z(T)} \sum_{s \in \mathbb{SS}(\aseq)} 
\left\{ \frac{E(s)}{RT^2} - \frac{1}{RT} \cdot \frac{\partial}{\partial T} E(s)
\right\} \cdot \exp(-E(s)/RT) \nonumber \\
&=& 
\sum_{s \in \mathbb{SS}(\aseq)} 
E(s) \cdot \frac{\exp(-E(s)/RT)}{Z(T)} - \nonumber \\
&& T \sum_{s \in \mathbb{SS}(\aseq)} 
\frac{\exp(-E(s)/RT)}{Z(T)} \cdot \frac{\partial}{\partial T} E(s) 
 \nonumber \\
&=& 
\langle E \rangle - T \cdot \langle \frac{\partial}{\partial T} E \rangle
\end{eqnarray}
Let {\em formal temperature} denote each occurrence of the temperature 
variable $T$ within the expression $RT$, while {\em table temperature}
denotes all other occurrences (i.e. table temperature refers to the
temperature-dependent Turner free energy parameters \cite{Turner.nar10}).
This will shortly be explained in greater detail.
From equation
(\ref{eqn:expEnergyStatPhysics}), it follows that
expected energy $\langle E \rangle$ is equal to $RT^2$ times
the derivative of $\ln Z(T)$ with respect to {\em formal temperature},
which latter we define to be the {\em formal temperature derivative} of
$\ln Z(T)$.

If we treat the energy $E(s)$ of structure $s$ as a constant (computed at
either the default temperature of $37^{\circ}$ C, or at a user-specified
temperature $T$), then the second term of 
equation (\ref{eqn:expEnergyStatPhysics2}) disappears, and we can approximate
$RT^2 \cdot \frac{\partial}{\partial T} \ln Z(T)$ by the finite difference
$RT^2 \cdot \frac{\ln Z(T+\Delta T) - \ln Z(T)}{\Delta T}$, where
for instance $\Delta T = 10^{-7}$. This requires a modification of
McCaskill's algorithm \cite{mcCaskill} for the partition function $Z(T)$,
where we distinguish between {\em formal temperature} and
{\em table temperature}.
Our software {\tt RNAentropy} implements such a modification, and thus
supports the formal temperature derivative (FTD) method of computing
thermodynamic structural entropy.

\begin{figure}[tbph]
\centering
\includegraphics[width=\textwidth]{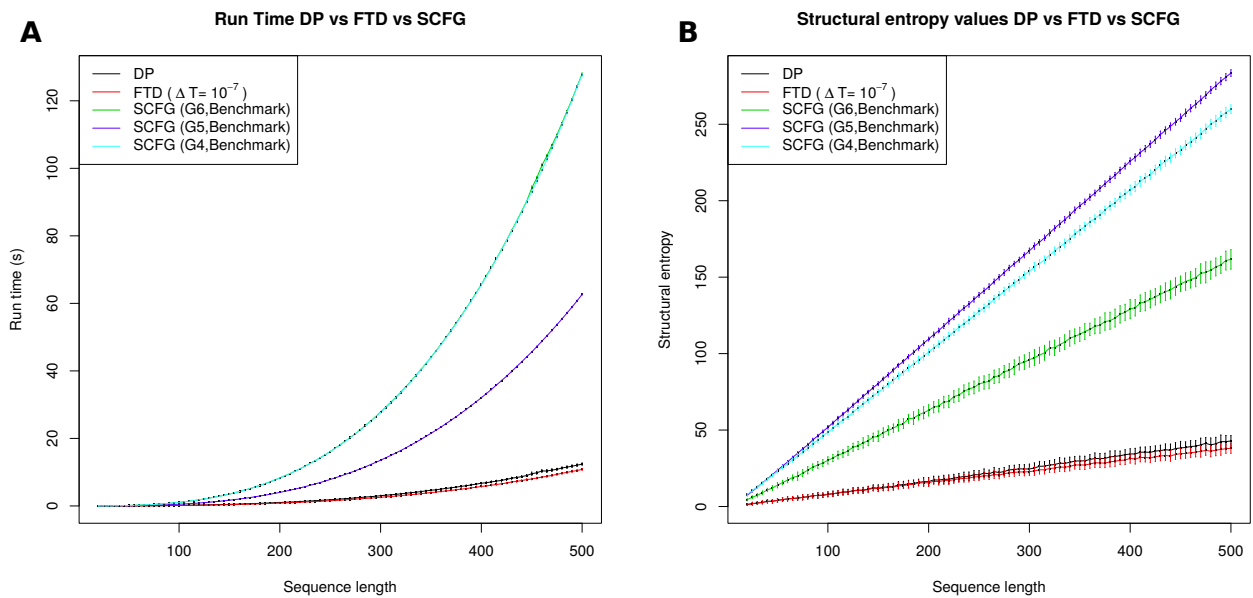}
\hskip 1cm
\caption{
{\em (A)} Average run times, with (tiny) error-bars of $\pm 1$ standard
deviation,  for each of the five methods DP, FTD ($\Delta T = 10^{-7}$),
SCFG(G6,Benchmark), SCFG(G4,Benchmark), and SCFG(G5,Benchmark).
Averages were determined for 100 random
RNA sequences of length $n$, each having expected compositional frequency of 
$0.25$ for A,C,G,U, where $n$ ranges from 20 to 500 with increments
of 5. 
Methods tested are as follows: (1) DP: dynamic programming computation of
expected energy $\langle E \rangle$ and partition function to yield
$H = \langle E \rangle/RT + \ln Z$, with Turner 2004 energy parameters.
(2) FTD: formal temperature derivative method which computes
$\langle E \rangle \approx RT^2 \cdot 
\frac{\ln Z(T+\Delta T) - \ln Z(T)}{\Delta T}$,
where the temperature increment $T+\Delta T$ is applied only to occurrences of
$T$ within the expression $RT$ -- i.e. {\em formal temperature}, as explained
in the text. Increment $\Delta T$ is $10^{-7}$, and 
Turner 2004 energy parameters are used.
(3) SCFG: computation of derivational entropy using the method of
\cite{Manzourolajdad.jtb13}, for the grammars G4, G5,
G6 with grammar rule probabilities from `Benchmark' data
(see \cite{Dowell.bb04,Manzourolajdad.jtb13}). 
SCFG executables and models downloaded from
{\tt http://rna-informatics.uga.edu/malmberg/}.
The methods, ordered from fastest to slowest, are as follows:
FTD, DP, G5, G6, G4, where FTD and DP are approximately equally fast,
while the slowest methods, G6 and G4, have almost identical run times.
DP and FTD are an order of magnitude faster than G6.
{\em (B)}
Average entropy values, with error bars of $\pm 1$ standard deviation, 
computed by the methods DP, FTD ($\Delta T = 10^{-7}$),
SCFG(G4,Benchmark), SCFG(G5,Benchmark), and SCFG(G6,Benchmark) 
for the same data set as in the left panel. The methods, ordered from
those returning smallest entropy values to largest, are as follows:
FTD, DP, G6, G4, G5.  FTD and DP return essentially identical values,
with a small deviation for larger sequences due to the finite approximation
of the formal temperature derivative.  
}
\label{fig:runtimes}
\end{figure}

Note that the function $\ln Z(T)$ is decreasing and concave down, so
barring numerical precision errors,
the finite difference $\frac{\ln Z(T+\Delta T) - \ln Z(T)}{\Delta T}$ 
is negative and slightly larger in absolute value than the
formal temperature derivative $\frac{\partial}{\partial T} \ln Z(T)$.
From equation (\ref{eqn:eqnShannonEntropyDef}), structural entropy
$H$ is equal to $\langle E \rangle/RT + \ln Z$ and so there will be a
small numerical deviation between the value of $H$, computed by the
FTD (formal temperature derivative) 
method currently described, and the exact value of $H$ computed by
the DP (dynamic programming) method, described in
Section~\ref{section:entropyDP}. In particular, entropy values computed
by FTD should be slightly smaller than those computed by DP, where
the discrepancy will be visible only for large sequence length. This
is indeed observed in Fig.~\ref{fig:runtimes}B and in data not shown.

%Figure 1
%end Figure 1

We now show that the expression,
$\langle \frac{\partial}{\partial T} E(s) \rangle$, occurring as the
second term in the last line of
equation~(\ref{eqn:expEnergyStatPhysics}), is equal to
$-T \cdot \langle S_t \rangle$ where $\langle S_t \rangle$  denotes the
expected change in entropy using the Turner parameters \cite{Turner.nar10}, 
determined as follows.  From statistical physics, the
free energy $E(s)$ of a secondary structure $s$ satisfies
\begin{eqnarray}
\label{eqn:mainThermo}
E(s) &=& H_t(s) - T\cdot S_t(s)
\end{eqnarray}
where $H_t(s)$ [resp. $S_t(s)$] denotes change in enthalpy  
[resp. entropy] from the empty structure to structure $s$ using the
Turner parameters.
The term $S_t$ measures the entropic loss due to stacked base pairs, 
hairpins, bulges, internal loops and multiloops using parameters obtained
from least-squares fitting of UV absorption data.
In the Turner energy  model, entropy $S_t$ and enthalpy $H_t$ are 
assumed to be independent of temperature, so it follows from
equation (\ref{eqn:mainThermo}) that
$\frac{\partial}{\partial T} E(s) = -S_t$, and hence
\begin{eqnarray}
\label{eqn:expEnergyStatPhysics2}
\langle E \rangle &=& RT^2 \frac{\partial}{\partial T} \ln Z(T) + T \cdot
\langle S_t \rangle
\end{eqnarray}
To compute $S_t(s)$ for a given secondary structure $s$ of an RNA sequence
$\aseq$, determine the the free energy $E(s,37)$ [resp. $E(s,38)$] of structure
$s$ at 37$^{\circ}$ C [resp. 38$^{\circ}$ C] by using
Vienna RNA Package {\tt RNAeval} \cite{Lorenz.amb11}; it then follows from
equation (\ref{eqn:mainThermo}) that
$S_t(s) = E(s,37)-E(s,38)$.  Throughout this paragraph, the
reader should not confuse the notion of
{\em conformational entropy} from equation(\ref{eqn:conformationalEntropy}), 
which is always non-zero and is computed by the novel algorithms described in
this paper, with the notion of {\em Turner change of entropy}
$S_t(s)$ of secondary structure $s$, which is always negative due to entropic
loss in going from the empty structure to a fixed structure $s$. 
Nor should the reader confuse
the notion of {\em structural entropy}, denoted by $H$ and defined in
equation (\ref{eqn:shannonEntropy}), with {\em Turner change of enthalpy}
$H_t(s)$ of secondary structure $s$.

%For example, let
%$\aseq$ denote the 78 nt arginyl-transfer RNA 
%from {\em Aeropyrum pernix} with tRNAdb accession code
%tdbR00000589 \cite{Juhling.nar09}. If $s$ is its structure of $\aseq$
%given in tRNAdb, then 
%%$S(\aseq,s) = -0.80999755859$, 
%$S(\aseq,s) = -0.81$, while if $s$ is the
%secondary structure consisting only of the GC base pair $(1,77)$ then
%%$S(\aseq,s) = -0.050000190735$. 
%$S(\aseq,s) = -0.05$. 
%(Note that $S(\aseq,s)$ actually represents change
%in entropy, i.e. entropy of $s$ minus entropy of the empty structure;
%however, using the notation $\Delta S$ would needlessly complicate 
%our current notation.)
%Since $S(\aseq,s)$ is a small negative value, it follows from
%equation (\ref{eqn:expEnergyStatPhysics2}) that
%$\langle E \rangle \approx RT^2 \frac{\partial}{\partial T} \ln Z(T)$, 
%although $\langle E \rangle$ is slightly larger than
%$RT^2 \frac{\partial}{\partial T} \ln Z(T)$.  This is illustrated in
%the right panel of Fig.~\ref{fig:runtimes}.

\subsection*{Entropy by dynamic programming}
\label{section:entropyDP}

Throughout this section, ${\bf a} = a_1,\cdots,a_n$ denotes an arbitrary
but fixed RNA sequence.  Below, we give recursions
for $Q(a)$, defined by $Q(a) = \sum_{s} BF(s) \cdot E(s)$, where the sum is 
taken over all secondary structures $s$ of RNA sequence ${\bf a}$,
$E(s)$ is the free energy of $s$, using the Turner 2004 parameters,
$BF(s)= \exp(-E(s)/RT)$ is the Boltzmann factor of structure $s$, where
$R$ is the universal gas constant and $T$ the temperature in Kelvin.

Recursions are also given for the 
partition function $Z(a) = \sum_s \exp(-E(s)/RT)$, where the sum is taken
over all secondary structures of ${\bf a}$. It follows that
the expected energy
\[
\langle E \rangle = \sum_{s} \frac{BF(s) \cdot E(s)}{Z} = \frac{Q(a)}{Z(a)}
\]
For $1\leq i\leq j \leq n$, the collection of
all secondary structures of $\aseq[i,j] = a_i,\ldots,a_j$ is denoted 
$\mathbb{SS}[i,j]$. In contrast,
if $s$ is a secondary structure of $a_1,\ldots,a_n$, then $s[i,j]$ is the
{\em restriction} of $s$ to the interval $[i,j]$, defined by
$s[i,j] = \{ (x,y) : i \leq x \leq y \leq j, (x,y) \in s$.

\subsection*{Initial steps}

For notational convenience, we define $Q_{i,i-1}=0$ and $Z_{i,i-1}=1$.
If $i \leq j < i+4$,  then for any secondary structure $s$, the restriction
$s[i,j]$ is the empty structure, denoted by $j-i+1$ dots with zero energy,
and so $Q_{i,j} = 0$.
As well, the only secondary structure on $[i,j]$ is the empty structure,
so $Z_{i,j}=1$. 

Now assume that $i+4 \leq j$. Since
\[
Q_{i,j}= \sum_{\substack{s \in \mathbb{SS}[i,j]\\ \text{$j$ unpaired in $s$}}} 
BF(s) E(s) + \sum_{k=i}^{j-4} 
\sum_{\substack{s \in \mathbb{SS}[i,j]\\ \text{$(k,j) \in s$}}} 
BF(s) E(s).
\]
we treat each sum in a separate case. Let $bp(k,j)$ be a boolean valued
function with the value $1$ if $k$ can base-pair with $j$; i.e. 
$a_ka_j \in \{ AU,UA,CG,GC,GU,UG \}$. For secondary structure 
$s\in \mathbb{SS}[i,j]$,
let $bp(k,j,s)$ be a boolean function with value $1$ if it is possible to
add the base pair $(k,j)$ to $s$ and obtain a valid secondary structure; i.e.
without creating a base triple or pseudoknot.
\medskip

\noindent
{\sc Case 1:} $j$ is unpaired in  $[i,j]$. For $s \in \mathbb{SS}[i,j]$ in which
$j$ is unpaired, $s=s[i,j-1]$, $BF(s)=BF(s[i,j-1])$, and $E(s) = E(s[i,j-1])$.
The contribution to
$Q_{i,j}$ in this case is given by $Q_{i,j-1}$.
\medskip

\noindent
{\sc Case 2:} $j$ is paired in  $[i,j]$.  The contribution to
$Q_{i,j}$ in this case is given by
\begin{eqnarray*}
Q_{i,j} &+=& \sum_{k=i}^{j-4} \sum_{\substack{s \in \mathbb{SS}[i,j]\\ \text{$(k,j) \in s$}}} 
BF(s) E(s) 
= \sum_{k=i}^{j-4} \sum_{\substack{s \in \mathbb{SS}[i,j]\\ 
\text{$(k,j) \in s$}}} 
BF(s) \left[ E(s[i,k-1]) + E(s[k,j]) \right]\\
&=& \sum_{k=i}^{j-4} bp(k,j) \cdot \left\{ 
\sum_{\substack{s_1 \in \mathbb{SS}[i,k-1]\\ \text{~}}} 
\sum_{\substack{s_2 \in \mathbb{SS}[k,j]\\ \text{$(k,j) \in s_2$}}} 
BF(s_1)\cdot BF(s_2)  \left[ E(s_1) + E(s_2) \right] \right\}\\ 
&=& \sum_{k=i}^{j-4}  bp(k,j) \cdot \left\{
\sum_{\substack{s_1 \in \mathbb{SS}[i,k-1]\\ \text{~}}} BF(s_1) E(s_1) 
\sum_{\substack{s_2 \in \mathbb{SS}[k,j]\\ 
\text{$(k,j)\in s_2$}}} BF(s_2) + \right. \\
&&\left. \sum_{\substack{s_1 \in \mathbb{SS}[i,k-1]\\ \text{~}}} BF(s_1) 
\sum_{\substack{s_2 \in \mathbb{SS}[k,j]\\ \text{$(k,j)\in s_2$}}} BF(s_2) E(s_2) 
\right\}  \\
&=& \sum_{k=i}^{j-4}  bp(k,j) \cdot \left\{
Q_{i,k-1} \cdot ZB_{k,j} +
Z_{i,k-1} \cdot QB_{k,j} \right\}. 
\end{eqnarray*}
Putting together the contributions from both cases, we have
\begin{eqnarray}
\label{eqn:nussTurnerExpNhbors}
Q_{i,j} &=& Q_{i,j-1} + \sum_{k=i}^{j-4} bp(k,j) \left[
Q_{i,k-1} ZB_{k,j} +
Z_{i,k-1} QB_{k,j}  \right].
\end{eqnarray}

\subsection*{Recursions for the Turner nearest neighbor energy model}
\label{section:turner}

In the nearest neighbor energy model 
\cite{zukerMathewsTurner:Guide,Turner.nar10}, free energies are
defined not for base pairs, but rather for {\em loops} in the loop
decomposition of a secondary structure. In particular, there are
stabilizing, negative free energies for stacked base pairs and
destabilizing, positive free energies for hairpins, bulges, internal loops,
and multiloops. 

In this section, free energy parameters for base stacking and loops
are from the Turner 2004 energy model \cite{Turner.nar10}.
As in the previous subsection, $Q,Z$ are defined, but now with respect 
to the Turner model.
\begin{eqnarray}
\label{eqn:Qij_defTurner}
Q_{i,j} &=& \sum_{s \in \mathbb{SS}[i,j]} E(s) \cdot \exp(-E(s)/RT)\\
Z_{i,j} &=& \sum_{s \in \mathbb{SS}[i,j]} \exp(-E(s)/RT). \nonumber
\end{eqnarray}
It follows that
$Z=Z_{1,n}$ is the partition function for secondary structures
(the Boltzmann weighted counting of all structures of $\aseq$)
and
\begin{eqnarray}
\label{eqn:NxBoltzmann}
\langle E(s) \rangle =
\frac{Q_{1,n}}{Z_{1,n}} = \sum_{s \in \mathbb{SS}[1,n]} p(s) \cdot E(s) =
\sum_{s \in \mathbb{SS}[1,n]} E(s) \cdot \frac{\exp(-E(s)/RT)}{Z}.
\end{eqnarray}
To complete the derivation of recursions, 
we must define $QB_{i,j}$ and $ZB_{i,j}$ for the Turner model.

To provide a self-contained treatment, 
we recall McCaskill's algorithm \cite{mcCaskill}, which efficiently computes
the partition function.
For RNA nucleotide sequence $\aseq = \aseq_1,\ldots,\aseq_n$, let
$H(i,j)$ denote the free energy of a hairpin closed by
base pair $(i,j)$, while
$IL(i,j,i',j')$ denotes the free energy of an {\em internal loop}
enclosed by the base pairs $(i,j)$ and $(i',j')$, where $i<i'<j'<j$.
Internal loops comprise the cases of
stacked base pairs, left/right bulges and proper internal loops.
The free energy for a multiloop containing 
$N_b$ base pairs and $N_u$ unpaired bases 
is given by the affine approximation $a+bN_b+cN_u$.
\medskip

%\begin{definition}[Partition function $Z$ and related function $Q$]
\noindent
{\sc Definition 1:} [Partition function $Z$ and related function $Q$]
\hfill\break
%\label{def:partitionFunctionDefMcCaskill}
\begin{itemize}
\item
$Z_{i,j} = \sum_{s} \exp(-E(s)/RT)$ where the sum is taken over
all structures $s \in \mathbb{SS}[i,j]$.
\item
$ZB_{i,j} = \sum_{s}  \exp(-E(s)/RT)$ where the sum is taken over
all structures $s \in \mathbb{SS}[i,j]$ which contain the base pair $(i,j)$.
\item
$ZM_{i,j} = \sum_{s} \exp(-E(s)/RT)$ where the sum is taken over
all structures $s \in \mathbb{SS}[i,j]$ which are contained
within an enclosing multiloop having {\em at least} one component.
\item
$ZM1_{i,j} = \sum_{s}  \exp(-E(s)/RT)$ where the sum is taken over
all structures $s \in \/Q[i,j]$ which are contained
within an enclosing multiloop having {\em exactly} one component.
Moreover, it is {\em required} that $(i,r)$ is a base pair of $x$,
for some $i<r\leq j$.
\item
$Q_{i,j} = \sum_{s} E(s) \cdot \exp(-E(s)/RT)$ where the sum is taken over
all structures $s \in ss[i,j]$.
\item
$QB_{i,j} = \sum_{s} E(s) \cdot \exp(-E(s)/RT)$ where the sum is taken over
all structures $s \in ss[i,j]$ which contain the base pair $(i,j)$.
\item
$QM_{i,j} = \sum_{s} E(s) \cdot \exp(-E(s)/RT)$ where the sum is taken over
all structures $s \in ss[i,j]$ which are contained
within an enclosing multiloop having {\em at least} one component.
\item
$QM1_{i,j} = \sum_{s}  E(s) \cdot \exp(-E(s)/RT)$ where the sum is taken over
all structures $s \in ss[i,j]$ which are contained
within an enclosing multiloop having {\em exactly} one component.
Moreover, it is {\em required} that $(i,r)$ is a base pair of $s$,
for some $i<r\leq j$.
\end{itemize}
%\end{definition}
\medskip

For $j-i \in \{0,1,2,3\}$, $Z(i,j)=1$, since the empty structure is the only
possible secondary structure. For $j-i>\theta=3$, we have
\begin{eqnarray*}
Z_{i,j} &= &Z_{i,j-1} + ZB_{i,j} + 
 \sum_{r=i+1}^{j-4} Z_{i,r-1} \cdot ZB_{r,j} \\
ZB_{i,j} &= &\exp(-HP(i,j)/RT) + 
\displaystyle\sum_{i \leq \ell \leq r \leq j}
\exp(-IL(i,j,\ell,r)/RT)\cdot ZB_{{\ell},r} +\\
&& \exp(-(a+b)/RT) \cdot \left( \sum_{r=i+1}^{j-\theta-2} ZM_{i+1,r-1}
\cdot ZM1_{r,j-1} \right)\\
ZM1_{i,j} &= &
\displaystyle\sum_{r=i+\theta+1}^j ZB_{i,r} \cdot
\exp(-c(j-r)/RT) \\
ZM_{i,j} &= &
\displaystyle\sum_{r=i}^{j-\theta-1}  ZM1_{r,j} \cdot
\exp(-(b+c(r-i))/RT)  + \\
&&\displaystyle\sum_{r=i+\theta+2}^{j-\theta-1}  ZM_{i,r-1} \cdot
ZM1_{r,j} \cdot \exp(-b/RT) . 
\end{eqnarray*}
\medskip

\noindent
{\sc Base Case:}
For $j-i \in \{ -1,0,1,2,3\}$, 
$Q_{i,j}= QB_{i,j}=0$, 
$Z_{i,j}=1$, $ZB_{i,j}= ZM_{i,j}= ZM1_{i,j}=0$. 
\medskip

\noindent
{\sc Inductive Case:} Assume that $j-i > 3$.  
\medskip

\noindent
{\sc Case A:} $(i,j)$ closes a hairpin.
\medskip

In this case, the contribution to $QB_{i,j}$ is given by 
\begin{eqnarray*}
A_{i,j} &=& \exp \left( - \frac{H(i,j)}{RT} \right)\cdot
H(i,j)
\end{eqnarray*}
\medskip

\noindent
{\sc Case B:} $(i,j)$ closes a stacked base pair, bulge or internal loop,
whose other closing base pair is $(\ell,r)$, where $i<\ell<r<j$.
\medskip

In this case, the contribution to $QB_{i,j}$ is given by the following
\begin{eqnarray*}
B_{i,j} &=& 
\sum_{\ell=i+1}^{\min(i+30,j-5)}
%\sum_{r=j-1}^{\max(j-30,i+5)}
\sum_{r=j-1}^{\max(\ell+4,j-(30-(\ell-i)))}
\sum_{\substack{s \in ss[\ell,r]\\ \text{$(\ell,r)\in s$}}}
\exp \left( - \frac{IL(i,j,\ell,r)}{RT} \right)\\
&&
\cdot BF(s) 
\cdot \left[
IL(i,j,\ell,r) + E(s) \right]\\\\
&=&
\sum_{\ell=i+1}^{\min(i+31,j-5)} 
%\sum_{r=j-1}^{\max(j-31,i+5)}
\sum_{r=j-1}^{\max(\ell+4,j-(30-(\ell-i)))}
\exp \left( - \frac{IL(i,j,\ell,r)}{RT} \right)\cdot IL(i,j,\ell,r)\\
&& \cdot 
ZB(\ell,r) + 
\exp \left( - \frac{IL(i,j,\ell,r)}{RT} \right)\cdot QB(\ell,r) .
\end{eqnarray*}
In the summation notation $\displaystyle\sum_{i=a}^b$, 
if upper bound $b$ is smaller than lower bound $a$, then we intend
a loop of the form: FOR $i=b$ downto $a$.
\medskip

\noindent
{\sc Case C:} $(i,j)$ closes a multiloop.
\medskip

In this case, the contribution to $QB_{i,j}$ is given by the following
\begin{eqnarray*}
C_{i,j} &=& 
\sum_{\substack{s \in ss[i,j], (i,j)\in s\\ \text{$(i,j)$ closes a multiloop}}} BF(s)E(s)\\ 
&=& 
%\sum_{r=i+5}^{j-5}
\sum_{r=i+6}^{j-5}
\exp \left( - \frac{a+b}{RT} \right)\cdot
\sum_{\substack{s_1 \in ss[i+1,r-1],s_2 \in ss[r,j-1]\\ 
\text{$r$ base-paired in $s_2$}}} BF(s_1) \cdot BF(s_2) \cdot \\
&&\left[a+b+E(s_1)+E(s_2) \right]\\
&=& 
%\sum_{r=i+5}^{j-5}
\sum_{r=i+6}^{j-5}
\exp \left( - \frac{a+b}{RT} \right)\cdot
\sum_{\substack{s_1 \in ss[i+1,r-1],s_2 \in ss[r,j-1]\\ 
\text{$r$ base-paired in $s_2$}}} BF(s_1) \cdot BF(s_2) \cdot \left[
a+b\right] +\\
&&
%\sum_{r=i+5}^{j-5}
\sum_{r=i+6}^{j-5}
\exp \left( - \frac{a+b}{RT} \right)\cdot
\sum_{\substack{s_1 \in ss[i+1,r-1]\\ \text{~}}} BF(s_1)  \cdot E(s_1)
\sum_{\substack{s_2 \in ss[r,j-1]\\ \text{$r$ base-paired in $s_2$}}} 
BF(s_2) + \\
&& 
%\sum_{r=i+5}^{j-5}
\sum_{r=i+6}^{j-5}
\exp \left( - \frac{a+b}{RT} \right)\cdot
\sum_{\substack{s_1 \in ss[i+1,r-1]\\ \text{~}}} BF(s_1) 
\sum_{\substack{s_2 \in ss[r,j-1]\\ \text{$r$ base-paired in $s_2$}}} 
BF(s_2) \cdot E(s_2) \\
&=& 
%\sum_{r=i+5}^{j-5}
\sum_{r=i+6}^{j-5}
\exp \left( - \frac{a+b}{RT} \right)\cdot \left[
(a+b) \cdot ZM(i+1,r-1) \cdot ZM1(r,j-1) + \right. \\
&&
\left. QM(i+1,r-1) \cdot ZM1(r,j-1) +
ZM(i+1,r-1) \cdot QM1(r,j-1) \right] .\\
\end{eqnarray*}
Now $QB_{i,j} = A_{i,j}+B_{i,j}+C_{i,j}$. 
It nevertheless remains to define the recursions for $QM1_{i,j}$ and
$QM_{i,j}$.
These satisfy the following.
\begin{eqnarray*}
QM1_{i,j} &=&\sum_{k=i+4}^{j} 
\sum_{\substack{s \in ss[i,k]\\ \text{$(i,k) \in s$}}} 
\exp \left( - \frac{c(j-k)}{RT} \right)\cdot BF(s) \cdot \left[
c(j-i) + E(s) \right] \\
 &=&\sum_{k=i+4}^{j} 
\exp \left( - \frac{c(j-k)}{RT} \right)\cdot \left[
c(j-i) \cdot ZB(i,k)  + QB_{i,k}  \right].
\end{eqnarray*}

\begin{eqnarray*}
QM_{i,j} &=& QMA_{i,j} + QMB_{i,j}\\
QMA_{i,j} &=&
\sum_{r=i}^{j-\theta-1} 
\sum_{\substack{s \in ss[r,j]\\ \text{$r$ pairs in $[r,j]$}}}
\exp \left( -\frac{b+c(r-i)}{RT} \right)\cdot 
\exp \left( -\frac{E(s)}{RT} \right)\cdot \left[ b+c(r-i) + E(s) \right]\\
&=&
\sum_{r=i}^{j-\theta-1} 
\exp \left( -\frac{b+c(r-i)}{RT} \right)\cdot  
\left\{ ZM1(r,j) \cdot (b+c(r-i)) +
QM1(r,j) \right\}  \\
QMB_{i,j} &=&
\sum_{r=i+5}^{j-\theta-1} 
\sum_{\substack{s_1 \in ss[i,r-1]\\ \text{~}}}
\sum_{\substack{s_2 \in ss[r,j]\\ \text{$r$ pairs in $[r,j]$}}}
\exp \left( -\frac{b}{RT} \right)\cdot 
\exp \left( -\frac{E(s_1)}{RT} \right)\cdot \\ && 
\exp \left( -\frac{E(s_2)}{RT} \right)\cdot 
\left[ b + E(s_1) + E(s_2) \right]\\
&=&
\exp \left( -\frac{b}{RT} \right)\cdot 
\sum_{r=i+5}^{j-4} 
\left\{ b \cdot ZM(i,r-1) \cdot ZM1(r,j)  + \right. \\
&& 
\left. QM(i,r-1) \cdot ZM1(r,j) +  ZM(i,r-1) \cdot QM1(r,j) \right\} . \\
\end{eqnarray*}
This completes the derivation of the recursions for expected energy.

\section*{Results}
\label{section:results}

In this section, we describe a detailed comparison of our thermodynamic
entropy algorithms FTD and DP, both implemented in the publicly available
program {\tt RNAentropy}, with the algorithm of 
Manzourolajdad et al.  \cite{Manzourolajdad.jtb13} which computes the
derivational entropy for trained RNA stochastic context free grammars.
Subsequently, we show that by accounting for structural entropy, there
is an improvement in the correlation between hammerhead ribozyme cleavage
activity and total free energy, extending a result of Shao et al.
\cite{Shao.bb07}.

\subsection*{Comparison of structural entropy and derivational entropy}
\label{section:comparison}

Using random RNA, 960 seed alignment sequences from Rfam family RF00005, 
and a collection of 2450 sequences obtained by selecting the
first RNA from the seed alignment of each family from
the Rfam 11.0 database \cite{Burge.nar13}, 
%we show that:
%(1) both of our algorithms DP, FTD compute the same 
%structural entropy values with the approximately the same efficiency, 
%(2) DP and FTD are
%substantially faster than the SCFG method of \cite{Manzourolajdad.jtb13},  
%(3) derivational entropy values computed by the
%SCFG method of \cite{Manzourolajdad.jtb13} are much larger than thermodynamic
%structural entropy values of DP and FTD, and
%(4) the correlation between thermodynamic structural entropy values and
%derivational entropy values is poor to moderately weak.
we show the following.
\begin{enumerate}
\item The thermodynamic structural entropy algorithms DP, FTD compute the same 
structural entropy values with the same efficiency, although as sequence
length increases, FTD runs somewhat faster and returns slightly smaller values 
than does DP, since FTD uses a finite difference to approximate the derivative
of the logarithm of the partition function.
\item  DP and FTD appear to be an order of magnitude 
faster than the SCFG method of \cite{Manzourolajdad.jtb13}, which latter
requires two minutes for RNA sequences of length 500 that require only a
few seconds for DP and FTD.
\item Derivational entropy values computed by the
method of \cite{Manzourolajdad.jtb13} are much larger than thermodynamic
structural entropy values of DP and FTD,
ranging from about 4-8 times larger,depending on the SCFG chosen.
\item
The length-normalized
correlation between thermodynamic structural entropy values and
derivational entropy values is poor to moderately weak.
\end{enumerate}
Unless otherwise specified, throughout this paper,
FTD, DP and SCFG refer to the 
formal temperature derivative method (Algorithm 1, with Turner'04 parameters),
the dynamic programming method (Algorithm 2, with Turner'04 parameters), and
the stochastic context free grammar method of
\cite{Manzourolajdad.jtb13}. SCFG(G4), SCFG(G5), SCFG(G6)
respectively refer to the SCFG method of \cite{Manzourolajdad.jtb13} 
using the stochastic context free grammars
G4, G5, and G6.  Additionally, there are three 
different training sets for each grammar: Rfam5, Mixed80 and Benchmark --
see \cite{Manzourolajdad.jtb13} for explanations of the training sets.
Thus SCFG(G6,Benchmark) refers to derivational entropy, computed by
the algorithm of \cite{Manzourolajdad.jtb13}, using grammar G6
with training set Benchmark, etc.

Table~\ref{table:RF00005entropyRunTime} lists the
average values, plus or minus one standard deviation, for the
entropy values and run time (in seconds) for 960 transfer RNAs from
the seed alignment of family RF00005
from Rfam 11.0 \cite{Burge.nar13}.  Results for
five methods are presented: (1) the dynamic programming method of
this paper, using the Turner 2004 free energy parameters (DP),
(2) approximating the formal temperature derivative 
$\frac{\partial}{\partial T} \ln Z(T)$ by finite differences, and 
subsequently applying equations
(\ref{eqn:expEnergyStatPhysics}, \ref{eqn:eqnShannonEntropyDef}),
using Turner 2004 free energy parameters (FTD);
(3,4,5) using the program of \cite{Manzourolajdad.jtb13} 
respectively with the stochastic context free grammars
G4, G5, and G6 trained on the dataset `Rfam5'. 
For the 960 transfer RNAs from the Rfam database, this table shows that
entropy values computed by DP and FTD are four to eight times smaller than 
derivational entropy values returned by the program of 
\cite{Manzourolajdad.jtb13}, while DP and FTD run five to ten times faster 
than the program of \cite{Manzourolajdad.jtb13} -- run times and 
derivational entropy values heavily depend on the grammar chosen and the
training set used for production rule probabilities.

%Table 1
\begin{table}[!ht]
\caption{Average values for structural entropy and run time 
(in seconds) for the 960 transfer RNA sequences
from the seed alignment of Rfam family RF00005. Methods include:
DP: dynamic programming algorithm from our program {\tt RNAentropy}, 
using the Turner 2004 energy parameters; 
FTD ($\Delta T = 10^{-7}$): finite difference computation of 
$\langle E \rangle = RT^2 \cdot \frac{\ln Z(T+\Delta T) - \ln Z(T)}{\Delta T}$,
where formal and table temperature are {\em uncoupled}, and 
formal temperature increment is $10^{-7}$; 
SCFG(G4,Rfam5): SCFG method \cite{Manzourolajdad.jtb13} 
using grammar G4 with training dataset `Rfam5';
SCFG(G5,Rfam5): SCFG method 
using grammar G5 with training dataset `Rfam5';
SCFG(G6,Rfam5): SCFG method 
using grammar G6 with training dataset `Rfam5'.
FTD returns very similar values for temperature increments
$10^{-7} \leq \Delta T \leq 10^{-11}$; however, for smaller temperature
increments, there is a slight deviation due to numerical precision issues
-- for example, average
entropy of FTD with $\Delta T = 10^{-12}$ is
$5.238878 \pm 1.504748$, with similar run times as other FTD runs.
}
\begin{tabular}{|l|rr|}
\hline
Method & Entropy ($\mu \pm \sigma$) & Run Time ($\mu \pm \sigma$) \\
\hline
DP & $5.953 \pm 1.381$ & $0.074 \pm 0.017$ \\
FTD ($\Delta T = 10^{-7}$) & $5.532 \pm 1.342$ & $0.058 \pm 0.014$ \\
SCFG(G4,Rfam5) & $39.917 \pm 2.885$ & $0.437 \pm 0.096$ \\
SCFG(G5,Rfam5) & $40.682 \pm 3.053$ & 0.204$ \pm 0.046$ \\
SCFG(G6,Rfam5) & $21.207 \pm 2.412$ & 0.433$ \pm 0.096$ \\
\hline
\end{tabular}

\label{table:RF00005entropyRunTime}
\end{table}
%end Table 1

Table~\ref{table:RF00005entropyCorr} presents the
Pearson correlation for entropy values of 960 transfer RNAs from
the seed alignment of family RF00005 from the
database Rfam 11.0 \cite{Burge.nar13}.  The upper-triangular 
[resp. lower-triangular] entries are correlations for
{\em unnormalized} [resp. {\em length-normalized}] entropy values.
Entropy values were computed for the same methods as in 
Table~\ref{table:RF00005entropyRunTime}.
Since there is little
variation in sequence length for the transfer RNAs in the seed alignment of 
RF00005 (average length is  $73.41 \pm 5.13$), any correlation due to
sequence length is eliminated. The table shows the poor correlation
between SCFG structural entropy, as computed by each grammar, with
thermodynamic structural entropy.

%Table 2
\begin{table}[!ht]
\caption{Pearson correlation for entropy values of 960 transfer RNAs from
the seed alignment of family RF00005 from the
database Rfam 11.0 \cite{Burge.nar13}.  Upper-triangular entries are
for {\em unnormalized} entropy values, while lower-triangular entries
are for {\em length-normalized} entropy values.  Entropy values were computed
for the same methods described in Fig.~\ref{table:RF00005entropyRunTime};
in particular, all SCFGs were trained with RF00005, as described in
\cite{Manzourolajdad.jtb13}.
}
\begin{tabular}{|l|lllll|}
\hline
Norm $\backslash$ Unnorm & DP &  FTD ($\Delta T = 10^{-7}$) & G4 & G5 & G6 \\
\hline
DP & 1 & 0.905 & 0.294 & 0.256 & 0.451 \\
FTD ($\Delta T = 10^{-7}$) & 0.919 & 1 & 0.142 & 0.116 & 0.398 \\
SCFG(G4,Rfam5) & 0.314 & 0.301 & 1 & 0.969 & 0.666 \\
SCFG(G5,Rfam5) & 0.247 & 0.263 & 0.720 & 1 & 0.619 \\
SCFG(G6,Rfam5) & 0.428 & 0.458 & 0.541 & 0.462 & 1 \\
\hline
\end{tabular}
\label{table:RF00005entropyCorr}
\end{table}
%end Table 2

Table~\ref{table:ZscoreEntropyGiegerichConformationalSwitches}
presents the average positional entropy, length-normalized
structural entropy, and corresponding
Z-scores for a small collection of experimentally confirmed
conformational switches, collected by Giegerich et al.
\cite{giegierich:RNAswitchesPSB99}, and available on the
{\tt RNAentropy} web server.  There appears to be no clear
entropic signal for conformational switches, at least with respect to
this small collection of sequences.

%Table 3
\begin{table}[!ht]
%\begin{adjustwidth}{-2.25in}{0in}
\caption{Thermodynamic structural entropy, positional entropy,
and corresponding
Z-scores for a small collection of experimentally confirmed
conformational switches, collected in
\cite{giegierich:RNAswitchesPSB99} -- sequences available at 
the {\tt RNAentropy} web site.
For each sequence, the positional (resp. structural)
entropy $x$ was computed, along with the mean $\mu$ and standard deviation
$\sigma$ of 1000 dinucleotide shuffles of the sequence. The Z-score is
then $\frac{x-\mu}{\sigma}$. Dinucleotide shuffles  were computed,
using the Altschul-Erikson algorithm \cite{altschulErikson:dinucleotideShuffle}
as implemented in \cite{cloteFerreKranakisKrizanc:RNA05}. 
Pearson correlation between Z-scores for positional and structural 
entropy is $0.4103$.
}
\begin{tabular}{|l|rrrrr|}
\hline
RNA	&	Seq len	&	Pos Ent	&	Norm str ent	&	Z-score, pos ent	&	Z-score, str ent	\\
\hline
Spliced-Leader & 56 & 0.802 & 0.075 & 0.755 & -0.697 \\
Attenuator & 73 & 0.326 & 0.054 & -0.871 & -0.983 \\
MS2 & 73 & 0.076 & 0.061 & -1.660 & -1.366 \\
S15 & 74 & 0.191 & 0.079 & -2.242 & -0.734 \\
E coli dsrA & 85 & 0.331 & 0.096 & -0.557 & 1.444 \\
HDV ribozyme & 107 & 0.326 & 0.034 & -2.037 & -2.424 \\
Tetrahymena Group I intron & 108 & 0.515 & 0.076 & -1.062 & 0.434 \\
E. coli alpha operon mRNA & 130 & 0.251 & 0.059 & -1.448 & -1.865 \\
hok & 142 & 0.340 & 0.087 & 0.700 & 0.608 \\
3'-UTR of AMV RNA & 145 & 0.336 & 0.077 & -0.517 & -0.316 \\
T4 td gene intron & 163 & 0.542 & 0.042 & -1.129 & -2.365 \\
thiM-Leader & 165 & 0.515 & 0.085 & -1.660 & -0.474 \\
btuB & 202 & 0.830 & 0.092 & -0.691 & 0.362 \\
Sbox-metE & 247 & 0.237 & 0.097 & -1.350 & 0.727 \\
HIV-1 leader & 280 & 0.324 & 0.086 & -1.425 & 0.109 \\
B. subtilis ribD leader & 304 & 0.471 & 0.067 & -1.835 & -1.460 \\
B. subtilis ypaA leader & 342 & 0.428 & 0.076 & -1.659 & -0.184 \\
\hline
\end{tabular}

\label{table:ZscoreEntropyGiegerichConformationalSwitches}
%\end{adjustwidth}
\end{table}
%end Table 3

Table~\ref{table:entropyExpEnergyEnsDef} presents the number of sequences,
average {\em length-normalized} thermodynamic entropy, average entropy Z-score,
average {\em length-normalized} ensemble defect, and average Z-score for
for sequences in the seed alignment of several RNA familes from
the Rfam 11.0 \cite{Burge.nar13}, as well as the 
precursor microRNAs from the repository MIRBASE \cite{GriffithsJones.nar08}.
Average values are given, plus or minus one standard deviation. The
Z-score is defined as $\frac{x-\mu}{\sigma}$, where $x$ is the entropy
(resp. ensemble defect) of a given sequence, and $\mu$ (resp. $\sigma$
denotes the mean (resp. standard deviation) of corresponding values for
100 random sequences having the same dinucleotides, obtained by using 
the Altschul-Erikson dinucleotide shuffling algorithm 
\cite{altschulErikson:dinucleotideShuffle}. As shown by this table,
Rfam family members appear to have lower structural entropy as well
as ensemble defect than random RNA having the same dinucleotides, although
the family RF00005 of transfer RNAs shows an exception to this rule for
structural entropy. The most pronounced Z-scores for structural entropy
and ensemble defect are for precursor microRNAs, which have very stable 
stem-loop structures. These results are generally comparable, with the
exception of entropy Z-scores for RF00005, with results concerning
minimum free energy (MFE) Z-scores from
\cite{rivasEddySecStrAlone,cloteFerreKranakisKrizanc:RNA05}. Indeed, the
particularly low MFE Z-scores of precursor miRNAs is used as a feature in the
support vector machine {\tt miPred} to detect microRNAs \cite{Ng.b07}.

%Table 4
\begin{table}[!ht]
%\begin{adjustwidth}{-2.25in}{0in} % Comment out/remove adjustwidth environment if table fits in text column.
\caption{For several large families from
the Rfam 11.0 database \cite{Burge.nar13}, and for
MIRBASE precursor microRNA \cite{GriffithsJones.nar08}, the table presents
the number of sequences (seq),
length-normalized values of thermodynamic structural
entropy (H) and ensemble defect (ens def), and the corresponding
Z-scores for entropy and ensemble defect.
For each sequence from a given RNA family, 100 random sequences were
generated with the same dinucleotides, using the Altschul-Erikson
dinucleotide shuffling algorithm \cite{altschulErikson:dinucleotideShuffle}
as implemented in \cite{cloteFerreKranakisKrizanc:RNA05} -- in the case
of MIRBASE, only 10 random sequences were generated for each sequence. 
Subsequently,
Z-scores were computed as $\frac{x-\mu}{\sigma}$, where $x$ is the 
entropy (resp. ensemble defect)
of a given sequence, and $\mu$ (resp. $\sigma$) is the mean (standard deviation)
of 100 random sequences having the same dinucleotides.
}
\begin{tabular}{|l|r|rr|rr|}
\hline
RNA family	&	seq	&	H	&	Z-score, H	&	ens def	&	Z-score, ens def	\\
\hline
RF00001 & 712 & $0.071 \pm 0.016$ & $-0.354 \pm 1.056$ & $0.198 \pm 0.123$ & $-0.423 \pm 0.965$ \\
RF00004 & 208 & $0.068 \pm 0.014$ & $-1.425 \pm 1.018$ & $0.177 \pm 0.103$ & $-0.901 \pm 0.863$ \\
RF00005 & 960 & $0.081 \pm 0.019$ & $-0.049 \pm 0.949$ & $0.189 \pm 0.105$ & $-0.405 \pm 0.820$ \\
RF00167 & 133 & $0.077 \pm 0.020$ & $-0.606 \pm 1.111$ & $0.164 \pm 0.105$ & $-0.782 \pm 0.858$ \\
MIRBASE & 28645 & $0.056 \pm 0.018$ & $-1.791 \pm 1.491$ & $0.101 \pm 0.076$ & $-1.324 \pm 0.791$ \\
\hline
\end{tabular}
\label{table:entropyExpEnergyEnsDef}
\end{table}
%end Table 4

We now turn to the figures that support each of the four assertions made
at the beginning of Section~\ref{section:comparison}.
Fig.~\ref{fig:runtimes} shows the average run times and entropy values for
for DP, FTD ($\Delta T = 10^{-7}$),
and the SCFG method of \cite{Manzourolajdad.jtb13} using each of the
grammars G4, G5 and G6 with training data from the
set `Benchmark'.  According to benchmarking work
of \cite{Manzourolajdad.jtb13} and \cite{Rivas.r12}, the grammar G6 seems
somewhat better than G4 and G5. It is for this reason that we focus principally
on the grammar G6, which was first introduced in the
SCFG algorithm {\tt Pfold} for RNA secondary structure prediction --
see \cite{Sukosd.bb11}.
The left panel of Fig.~\ref{fig:runtimes}A depicts average run times
for DP, FTD, and SCFG methods, for 100 random
RNA sequences of length $n$, where $n$ ranges from 20 to 500 with an increment
of 5. This figure shows that FTD and DP run faster by an order of
magnitude than the SCFG methods -- indeed, for
length 500 RNAs, derivational entropy is computed in two minutes, while
thermodynamic structural entropy is computed in a few seconds.
The Fig.~\ref{fig:runtimes}B depicts the entropy values
computed by DP, FTD ($\Delta T = 10^{-7}$), and SCFG methods.
Note that for large RNA sequence length, entropy values
returned by FTD are slightly smaller than those returned by DP, in
agreement with the discussion in Section~\ref{section:statisticalPhysics}.
Entropy values for the grammar G5 are considerably larger than those
of FTD and DP, while entropy values for G4 and G6 are almost
identical and approximately twice the size of those from G5.

\begin{figure}[tbph]
\centering
\includegraphics[width=\textwidth]{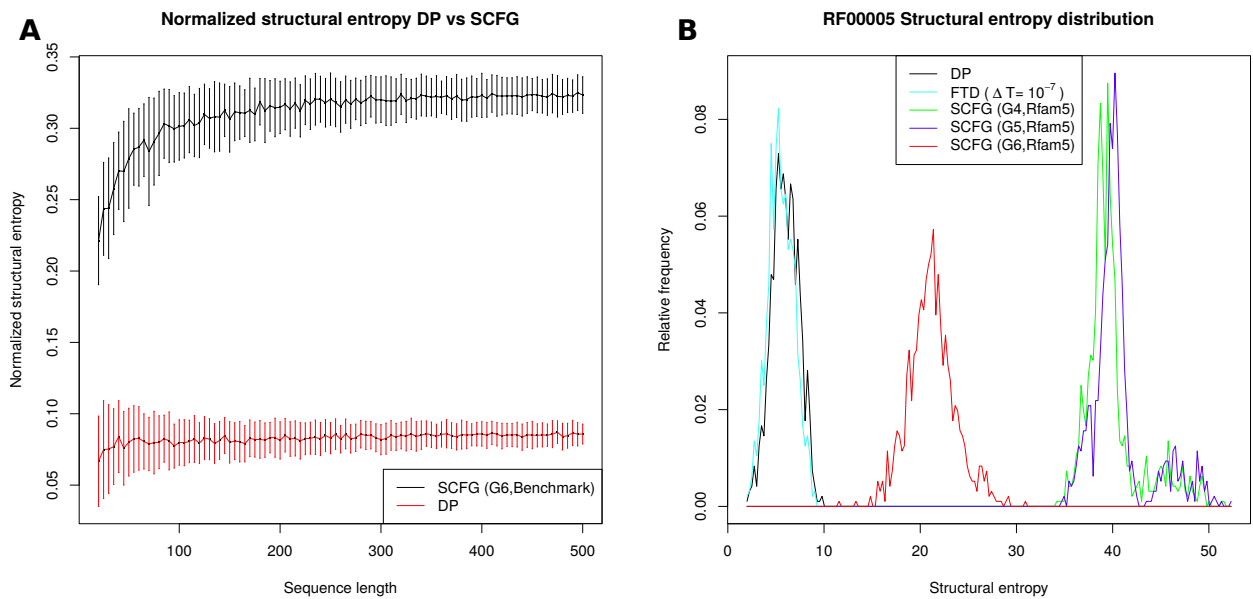}
\caption{
{\em (A)} The average of
length-normalized entropy values, as computed by DP and
SCFG (G6,Benchmark), using the same data as described in the
caption of Fig.~\ref{fig:runtimes}.
Using methods from algebraic combinatorics, it can be proven
that the length-normalized entropy for a homopolymer is asymptotically
constant. By numerical fitting, we find that SCFG values are roughly 
four times as large as DP values (approximate fitted value 3.78).
{\em (B)}
Relative frequency of entropy values for the 960
transfer RNA sequences in the seed alignment of RF00005 family from
Rfam 11.0 \cite{Burge.nar13}, as computed for each of 
the five methods DP, FTD ($\Delta T = 10^{-7}$),
SCFG(G4,Rfam5), SCFG(G5,Rfam5) and SCFG(G6,Rfam5).
See the caption from Fig.~\ref{fig:runtimes} for explanation of each
method, where in contrast to previous figures, the training set `Rfam5'
was used in place of `Benchmark'.
Average entropy values for RF00005 are given as follows.
FTD ($\Delta T = 10^{-7}$): $5.53 \pm 1.34$.
DP: $5.95 \pm 1.38$;
G4: $39.92 \pm 2.88$;
G5: $40.68 \pm 3.05$;
G6: $21.21 \pm 2.41$.
Note the bimodal distribution of 
entropy values computed with the SCFGs G4 and G5.
Relative frequency plot for 712 5S ribosomal RNAs from RF00001
is very similar (data not shown).
}
\label{fig:lengthNormEntropy}
\end{figure}

Fig.~\ref{fig:lengthNormEntropy}A presents graphs of
length-normalized entropy values, computed by DP and SCFG. Using methods
from algebraic combinatorics \cite{Lorenz.jcb08,Fusy.jmb14},
it is possible to prove that the length-normalized
asymptotic structural entropy is constant, as observed in this figure.
By numerical fitting, we find that the slope of 
the DP line is $0.087$, while 
that of G6 is $0.329$; i.e. SCFG entropy
values using the G6 grammar are 3.78 times those of DP entropy. 
This is supported by Table~\ref{table:RF00005entropyRunTime}, which 
suggests that G6 entropy values are 3.56 times larger than DP,
while G4 and G5 entropy values are 6.71 resp.  6.85
times larger than DP entropy values.
Fig.~\ref{fig:lengthNormEntropy}B 
depicts the relative frequency of structural entropy values for DP, 
FTD, and SCFG  methods for 960 transfer RNA sequences from the seed 
alignment of the Rfam 11.0 database \cite{Burge.nar13}.

%Figure 2
%end Figure 2

\begin{figure}[tbph]
\centering
%\includegraphics[width=0.45\textwidth]{figure3a}
%\qquad
%\includegraphics[width=0.45\textwidth]{figure3a}
\includegraphics[width=\textwidth]{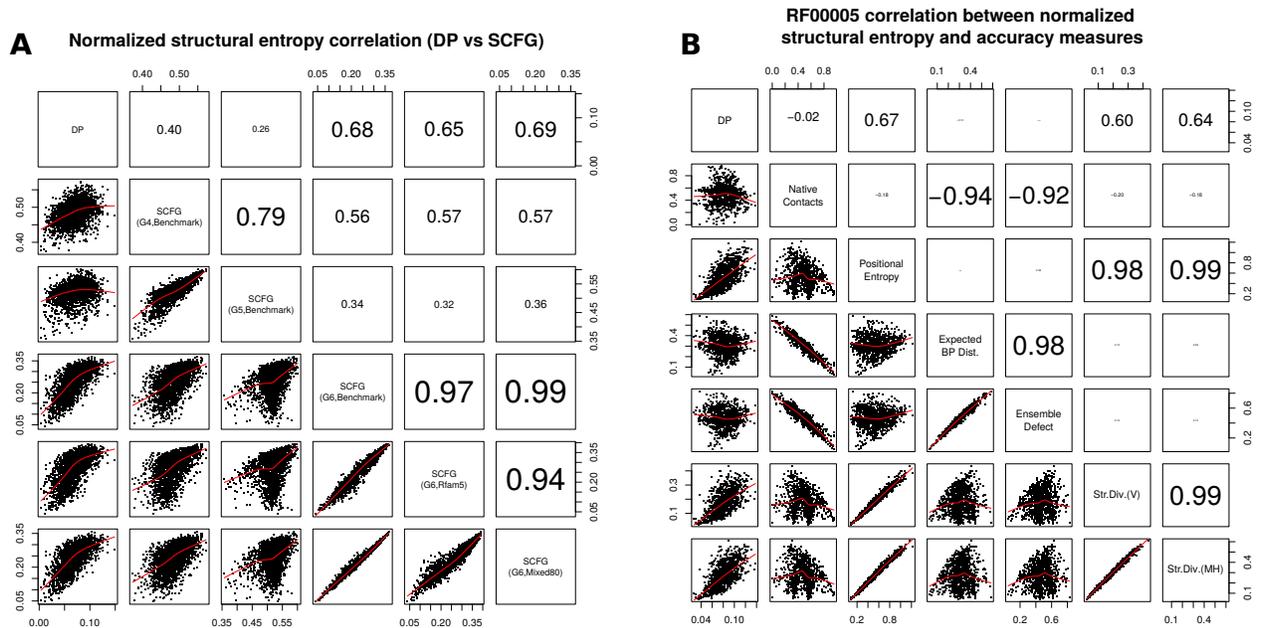}
\caption{
{\em (A)} Correlation between length-normalized
structural entropy values, as computed
by DP and five stochastic context free grammars:
grammars G4, G5 and G6 for the `Benchmark' training
set, and G6 for `Rfam5' and `Mixed80' training sets
(see \cite{Manzourolajdad.jtb13}). Low correlation is shown
between length-normalized thermodynamic structural and derivational entropies.
For the fixed grammar G6, very high
correlation is displayed between length-normalized entropy values for
each of the training sets `Benchmark', `Rfam5', `Mixed80' (similar results
for fixed grammars G4,G5 -- data not shown). Although
grammars G4 and G5 display a moderately high correlation
together, there is low correlation with length-normalized entropy values
determined by the grammar G6.
Benchmarking set consists of
the first sequence in the seed alignment from each family in the
database Rfam 11.0 \cite{Burge.nar13}. 
{\em (B)} Scatter plots and 
correlation between thermodynamic structural entropy
and several measures of {\em structural diversity},
computed from 960 tRNA sequences in the seed alignment of family
RF00005 from from the Rfam 11.0 database \cite{Burge.nar13}.
Correlation is computed between the following normalized values:
(1) DP: length-normalized
thermodynamic structural entropy computed by DP algorithm.
(2) Native Contacts: proportion of base pairs in the Rfam consensus structure
that appear in the low energy Boltzmann ensemble, defined by
$\sum_s p(s) \cdot \frac{|s \cap s_0|}{|s_0|}$, where $s_0$ is the
Rfam consensus structure.
(3) Positional Entropy: average positional entropy, defined by
$\sum_{i=1}^n H_2(i)/$, where $H_2(i)$
is defined by equation~(\ref{def:positionalStrEntropy}).
(4) Expected base pair distance: length-normalized value determined from
$\sum_s p(s) \cdot d_{\mbox{\tiny BP}}(s,s_0)$, where $s_0$ is the Rfam
consensus structure, which equals
$\sum_{1 \leq i<j \leq n}
I[ (i,j) \not\in s_0 ] \cdot p_{i,j} +
I[ (i,j) \in s_0 ] \cdot (1-p_{i,j})$
where $I$ denotes the indicator function -- see \cite{Dotu.nar15}.
(5) Ensemble defect: length-normalized value determined from
$n - \sum_{i \ne j} p^*_{i,j} \cdot I[ (i,j) \in s_0 ]
- \sum_{1 \leq i \leq n} p^*_{i,i}\cdot I[ \mbox{$i$ unpaired in $s_0$}]$,
where $I$ denotes the indicator function, and $p^*_{i,j}$ is defined
in equation (\ref{def:pijstar}) -- see \cite{RNAdesignNAR2004}.
(6) Str. Div. (V): Vienna structural diversity, output as 
{\em ensemble diversity} by {\tt RNAfold -p} \cite{Lorenz.amb11}.
(7) Str. Div. (MH): Morgan-Higgs structural diversity
\cite{morganHiggsBarrier}, defined in the text.
Positional entropy is moderately correlated with DP; ensemble defect
and expected base pair distance are highly correlated, and each
is moderately correlated with the proportion of native contacts.
Structural diversity (Vienna and Morgan-Higgs) are highly correlated with
positional entropy, but only (surprisingly) only moderately correlated with 
conformational entropy DP, in spite of the fact that all these measures
concern properties of the ensemble of structures.  Ensemble defect,
expected base pair distance and expected number of native contacts are all
highly correlated; this is unsurprising, since all measures concern 
the deviation of structures in the ensemble from the minimum free energy
structure.
Note that positional entropy is poorly correlated with the proportion of
native contacts, although Huynen et al. \cite{Huynen.jmb97} show
that base pairs in the MFE structure of 16S rRNA tend to belong to the
structure determined by comparative sequence analysis when the nucleotides
have low positional entropy.
}
\label{fig:correlationG6}
\end{figure}

Fig.~\ref{fig:correlationG6} presents scatter plots
and Pearson correlation of length-normalized entropy values 
and several notions of structural diversity that have been used
for RNA design \cite{Zadeh.jcc11,Dotu.nar15}. Values were computed
in this figure for a set of 2450 RNAs of various lengths, by selecting
the first sequence from the seed alignment of each family from 
the Rfam 11.0 database \cite{Burge.nar13}, after discarding a
few families having too few sequences.
Fig.~\ref{fig:correlationG6}A
depicts the Pearson correlation between length-normalized
structural entropy values, as computed
by DP, FTD, and the SCFG method using grammars G4, G5, G6.
Length-normalized derivational entropy values 
remain highly correlated, regardless of training set, but the correlation
of all SCFG methods is poor with DP. The Pearson correlation of 0.79 for
length-normalized entropy values obtained by G4 and G5 is high; however
the correlation with G6 drops to 0.56 (G4-G6) and 0.34 (G5-G6).
Fig.~\ref{fig:correlationG6}B 
depicts scatter plots and Pearson correlation for length-normalized
structural entropy, as computed by DP, and various notions of
structural diversity used in synthetic RNA design. (By minimizing values
such as the positional entropy,
structural entropy, ensemble defect, expected base pair distance, it
is more likely that computationally designed RNAs will fold into their
predicted structures when experimentally validated.)
Brief definitions of the notions of structural diversity that are compared
in Fig.~\ref{fig:correlationG6}B are given as follows.
{\em Native Contacts}: proportion of base pairs in the Rfam consensus structure
that appear in the low energy Boltzmann ensemble, defined by
$\sum_s p(s) \cdot \frac{|s \cap s_0|}{|s_0|}$, where $s_0$ is the
Rfam consensus structure.
{\em Positional Entropy}: average positional entropy 
$\sum_{i=1}^n H_2(i)/n$, where $H_2(i)$
is defined by equation~(\ref{def:positionalStrEntropy}).
{\em Expected base pair distance}: length-normalized value determined from
$\sum_s p(s) \cdot d_{\mbox{\small BP}}(s,s_0)$, where $s_0$ is the Rfam
consensus structure, computed by
$\sum_{1 \leq i<j \leq n}
I[ (i,j) \not\in s_0 ] \cdot p_{i,j} +
I[ (i,j) \in s_0 ] \cdot (1-p_{i,j})$
where $I$ denotes the indicator function -- see \cite{Dotu.nar15}.
{\em Ensemble defect}: length-normalized value determined from
$n - \sum_{i \ne j} p^*_{i,j} \cdot I[ (i,j) \in s_0 ]
- \sum_{1 \leq i \leq n} p^*_{i,i}\cdot I[ \mbox{$i$ unpaired in $s_0$}]$,
where $s_0$ is the Rfam consensus structure,
$I$ denotes the indicator function, and $p^*_{i,j}$ is defined
in equation (\ref{def:pijstar}).
{\em Vienna structural diversity}: Boltzmann average base pair distance
between each pair of structures in the ensemble, called 
{\em ensemble diversity} in the output of {\tt RNAfold -p} \cite{Lorenz.amb11},
formally defined by
$\sum_{i<j} p_{i,j}(1-p_{i,j}) + (1-p_{i,j})p_{i,j}$, where $p_{i,j}$ 
and output as {\em ensemble diversity} by {\tt RNAfold -p}.
{\em Morgan-Higgs structural diversity}: Boltzmann average Hamming distance
between each pair of structures in the ensemble, where a structure $s$ is
represented by an array where $s[i]=j$ if $(i,j)$ or $(j,i)$ is a base pair,
and otherwise $s[i]=i$, formally defined by
$n - \sum_{i,j} p^*_{i,j} \cdot p^*_{i,j}$.
Length-normalized DP entropy values are moderately highly correlated with
positional entropy, but not with the other measures. In synthetic design
of RNAs, it is our opinion that one should prioritize for experimental
validation those synthetically designed RNAs by consideration of
ensemble defect, structural entropy, etc., where the measures selected
are not highly correlated. From this standpoint, one might
use ensemble defect, structural entropy and proportion of native contacts
as suitable measures for synthetic RNA design -- see \cite{RNAdesignNAR2004}.

%Figure 3
%end Figure 3

\begin{figure}[tbph]
\centering
\hskip 1cm
\includegraphics[width=1.00\textwidth]{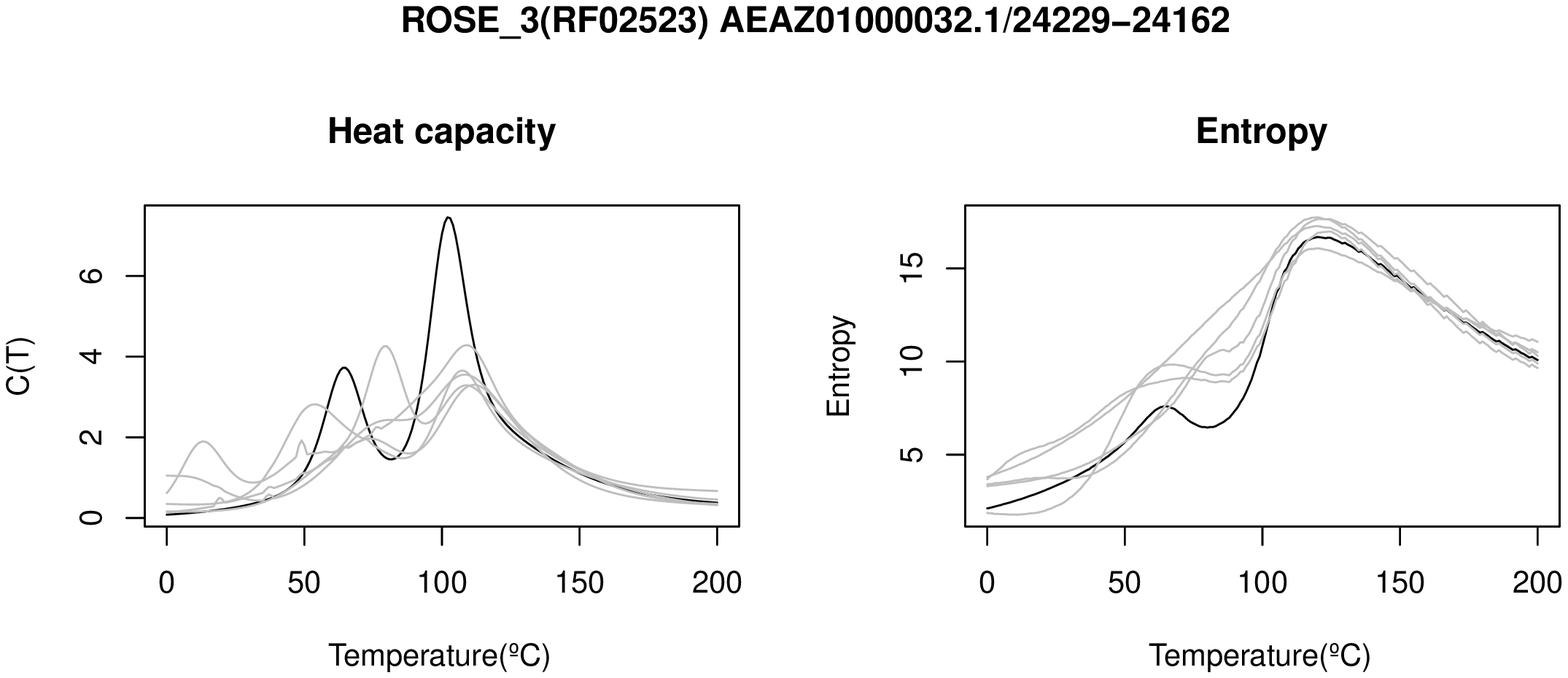}
\caption{Heat capacity (left) and thermodynamic structural entropy (right)
for a thermoswitch, or RNA thermometer, 
from the ROSE 3 family RF02523 from the Rfam 11.0 database 
\cite{Burge.nar13}, with EMBL accession code
AEAZ01000032.1/24229-24162. Lighter curves in the background
correspond to the heat capacity (left) and thermodynamic structural
entropy (right) of random RNAs having the same dinucleotides, obtained
by the implementation in \cite{cloteFerreKranakisKrizanc:RNA05}
of the Altschul-Erikson dinucleotide shuffle algorithm
\cite{altschulErikson:dinucleotideShuffle}.  Since structural entropy
$H = \langle E(T) \rangle/RT + \ln Z(T)$ and heat capacity
$C(T) = \frac{\partial}{\partial T} \langle E(T) \rangle$, the derivative
of entropy $H$ with respect to temperature closely follows the curve of
the heat capacity (data not shown). Heat capacity computed using
Vienna RNA Package {\tt RNAheat} \cite{Lorenz.amb11}, and entropy computed
by method DP.
}
\label{fig:heatCapacity}
\end{figure}

Fig.~\ref{fig:heatCapacity} displays the heat capacity and structural
entropy for a thermoswitch (also called RNA thermometer) from
the ROSE 3 family RF02523 from the 
Rfam 11.0 database \cite{Burge.nar13}, 
with EMBL accession code AEAZ 01000032.1/24229-24162.
The heat capacity, computed by Vienna RNA Package {\tt RNAheat},
presents two peaks, corresponding to two critical temperatures
$T_1,T_2$, where one of the two conformations of this thermoswitch is 
stable in the temperature range between $T_1$ and $T_2$.
The entropy plot also suggests the presence of a stable structure
in the temperature range between $T_1$ and $T_2$, since small entropy
values entail small diversity in the Boltzmann ensemble of structures.

%Figure 4
%end Figure 4

As shown in the tables and figures, the DP and FTD methods return
almost identical values and have very similar (fast) run times,
contrasted with the SCFG method, which is slow and whose values are
much larger than those of DP and FTD. For a sequence of length
500, SCFG(G6,Benchmark) takes 2 minutes, compared with a few seconds for
DP and FTD. Since
FTD approximates a derivative by a finite difference, one expects a small
discrepancy in the values of DP and FTD for
thermodynamic structural entropy. 
According to \cite{Manzourolajdad.jtb13},
the sensitivity and specificity of G4 and G6
grammars are ``significantly'' higher than that of the G5 grammar.
Since G6 is the underlying grammar of the {\tt Pfold} software,
for many of our comparisons, we compute derivational entropy using
grammar G6 with the `Benchmark' training set.
(In data not shown, we benchmarked all
nine combinations of grammars and training sets.)

\subsection*{Using {\tt RNAfold} to compute conformational entropy}
\label{section:negativeEntropyViennaPackage}

\begin{figure}[tbph]
\centering
%\includegraphics[width=0.45\textwidth]{figureViennaRuntime}
%\includegraphics[width=0.45\textwidth]{figure5a}
%\qquad
%\includegraphics[width=0.45\textwidth]{figureViennaStdev}
%\includegraphics[width=0.45\textwidth]{figure5b}
\includegraphics[width=\textwidth]{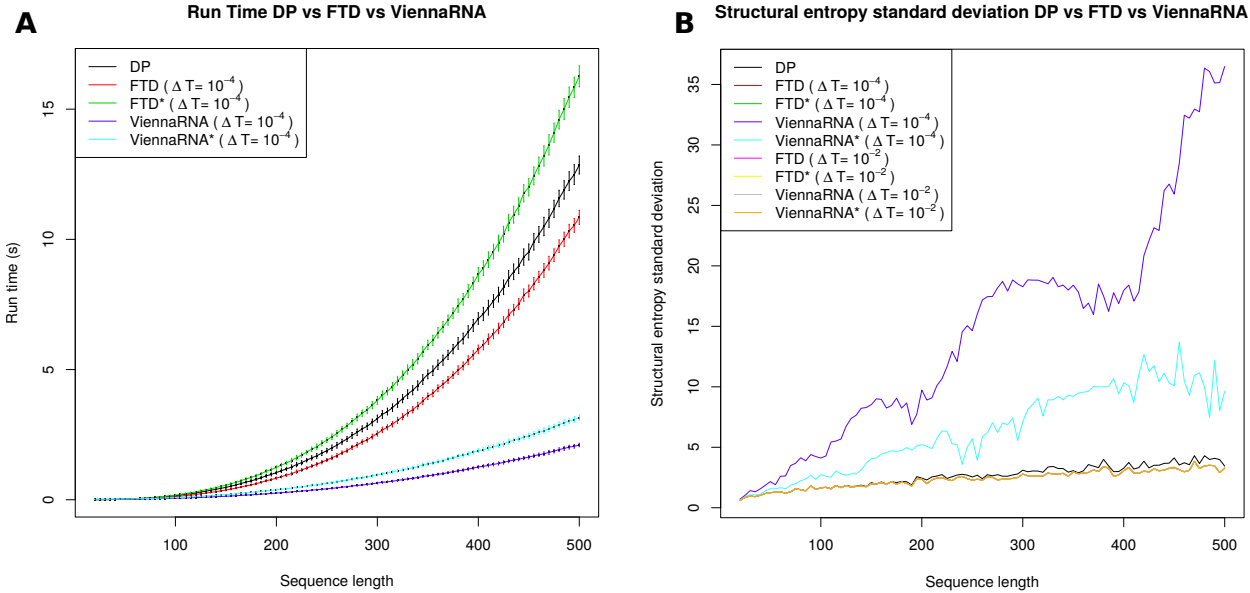}
\caption{
Average values for the run time and the entropy values for 100 random
RNA sequences of length $n$, each having expected compositional frequency of
$0.25$ for A,C,G,U, where $n$ ranges from 20 to 500 with increments
of 5 for conformational entropy.
{\em (A)} Average run times as a function of sequence length,
where error bars represent $\pm 1$ standard deviation.
Methods used: {\tt DP}, {\tt FTD}, 
{\tt FTD}$^*$, {\tt ViennaRNA}, {\tt ViennaRNA}$^*$. For random RNAs of
length 500 nt, {\tt Vienna RNA Package} is about three times faster than
our code.
{\em (B)} Standard deviation of the entropy values computed for 100 random
RNA, displayed as a function of sequence length. 
From top to bottom, the first three curves represent
uncentered {\tt ViennaRNA} with $\Delta T=10^{-4}$,
centered {\tt ViennaRNA}$^*$ with $\Delta T=10^{-4}$,
and DP. The bottom curve represents
centered {\tt FTD} with $\Delta T=10^{-4}$,
centered {\tt FTD}$^*$ with $\Delta T=10^{-2}$,
uncentered {\tt ViennaRNA} with $\Delta T=10^{-2}$,
centered {\tt ViennaRNA}$^*$ with $\Delta T=10^{-2}$.
The average entropy values
computed by {\tt FTD}, {\tt FTD}$^*$, {\tt ViennaRNA}, and 
{\tt ViennaRNA}$^*$ are indistinguishable and since {\tt FTD} values are
shown in the right panel of Fig.~\ref{fig:runtimes}, they are not shown
here.
}
\label{fig:viennaRuntimes}
\end{figure}

\begin{figure}[tbph]
\centering
%\includegraphics[width=0.45\textwidth]{figure6a}
%\qquad
%\includegraphics[width=0.45\textwidth]{figure6b}
\includegraphics[width=\textwidth]{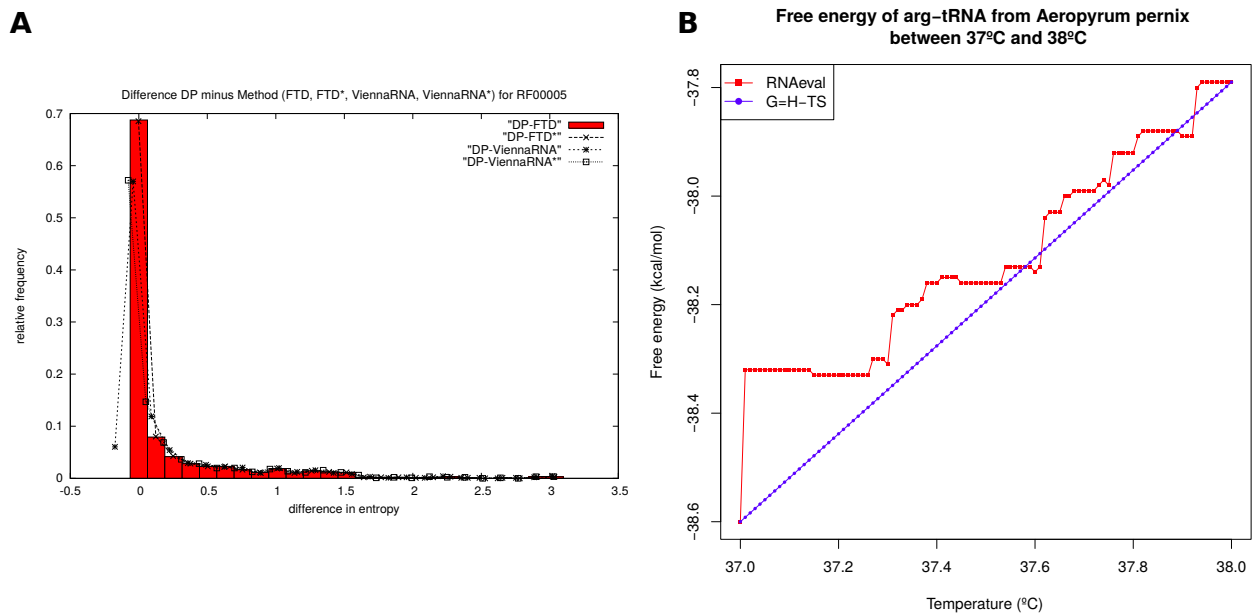}
\caption{{\em (A)}
Relative frequency of the difference in entropy values
for 960 transfer RNAs from
the RF00005 family of the Rfam 11.0 database.
(1) DP-FTD with average entropy difference
$0.2512  \pm 0.4935$ with maximum of 3.1622  and minimum of 0.
(2) DP-FTD$^*$ with average entropy difference
$0.2502 \pm 0.4934$ with maximum of 3.1602 and minimum of -0.0020.
(3) DP-ViennaRNA with average entropy difference
$0.2475 \pm 0.4975$ with maximum of 3.1520 and minimum of -0.1743.
(4) DP-ViennaRNA$^*$ with average entropy difference
$0.2494 \pm 0.4946$ with maximum of 3.1572 and minimum of -0.0777.
% Avg, stdev, max, min of absolute values of same differences:
%Mean:0.250611	StDev:0.493323	Max:3.160293	Min:0.000004
%Overflow  0.104167 %
%Mean:0.250611	StDev:0.493323	Max:3.160293	Min:0.000004
%Overflow  0.104167 %
%Mean:0.262836	StDev:0.489612	Max:3.152007	Min:0.000068
%Overflow  0.104167 %
%Mean:0.255943	StDev:0.491309	Max:3.157177	Min:0.000047
%Overflow  0.104167 %
%
It is noteworthy that FTD is {\em always} less than DP, 
FTD$^*$ exceeds DP by a tiny margin only rarely, while
ViennaRNA and ViennaRNA$^*$  more often exceed DP. Recall that the
average deviation DP-FTD increases with increasing sequence length, as shown
in the right panel of Figure~\ref{fig:runtimes}. The same is true 
for DP-FTD$^*$, DP-ViennaRNA, DP-ViennaRNA$^*$  (data not shown).
{\em (B)}
Free energy of arginyl-transfer RNA from 
{\em Aeropyrum pernix} with tRNAdb accession code tdbR00000589 
\cite{Juhling.nar09} for temperatures ranging from 37$^{\circ}$ C
to $38^{\circ}$ C in increments of $0.01$. The blue piecewise linear
curve was created using {\tt RNAeval -T} from the Vienna RNA Package
\cite{Lorenz.amb11}. The red linear curve was created by 
(1) calculating the entropy $S_t = G(37)-G(38)$ of the tRNA cloverleaf 
structure by subtracting the 
free energy at 38$^{\circ}$ C from the free energy at 37$^{\circ}$ C,
as determined using {\tt RNAeval -T}, (2) computing the
enthalpy $H_t = G(37)+(273.15 + 37) \cdot S_t$, and then
(3) computing the free energy at temperature $T$ by
$G(T) = H_t - T \cdot S_t$.  The jagged free energy curve is due to
the fact that Vienna RNA Package represents energies as integers
(multiples of 0.01 kcal/mol), so that loop energies jump at 
particular temperatures. 
}
\label{fig:RNAfoldEntropy}
\end{figure}

We have recently learned that newer versions of Vienna RNA Package
\cite{Lorenz.amb11} allow the user to modify the value $RT$ by using 
the flag {\tt --betaScale} (kindly pointed out by Ivo Hofacker).
It follows that {\tt RNAfold} can easily be used to compute conformational
entropy by using the FTD method. 
Let $T = 310.15$ be the absolute temperature corresponding to 
37$^{\circ}$ C, let $\Delta T = 0.01$, let $T_2 = {T+\Delta T} = 310.16$ 
and $T_1 = {T-\Delta T} = 310.14$. Define the scaling factors
$\beta_2 = \frac{T+ \Delta T}{T} = 1.0000322424633241$, and
$\beta_1 = \frac{T- \Delta T}{T} = 0.9999677575366759$. Run
{\tt RNAsubopt -p --betaScale $\beta_2$} to compute the ensemble free energy
$-R(T+\Delta T) \ln Z(T+\Delta T)$, and
{\tt RNAsubopt -p --betaScale $\beta_1$} to compute the ensemble free energy
$-R(T-\Delta T) \ln Z(T-\Delta T)$, where 
$Z(T+\Delta T)$ [resp. $Z(T-\Delta T)$] temporarily denotes the value of
the partition function where table temperature is 37$^{\circ}$ C (as usual),
and formal temperature is $T + \Delta T$ [resp. $T - \Delta T$] in Kelvin.
It follows that the {\em uncentered finite difference}
equation~(\ref{eqn:tonoExpectedEnergyApprox})
\begin{eqnarray}
\label{eqn:tonoExpectedEnergyApprox}
R T^2 \cdot \frac{ \ln Z(T + \Delta T) - \ln Z(T)}{\Delta T}
\end{eqnarray}
as well as the {\em centered finite difference}
\begin{eqnarray}
\label{eqn:ivoExpectedEnergyApprox}
R T^2 \cdot \frac{ \ln Z(T + \Delta T) - \ln Z(T - \Delta T)}{ 2 \Delta T}
\end{eqnarray}
both provide good approximations for the expected energy $\langle E \rangle$.
Now run {\tt RNAsubopt -p } to compute the ensemble free energy
$G = -RT \ln Z$ where table and formal temperature are (as usual) 310.15 in
Kelvin, and so compute the entropy
\begin{eqnarray}
\label{eqn:entropyFTD}
H &= & \frac{ \langle E \rangle - G}{RT}.
\end{eqnarray}
Let {\tt ViennaRNA} [resp. {\tt ViennaRNA}$^*$] denote the entropy  computation
just described, where expected energy is approximated by the
uncentered equation~(\ref{eqn:tonoExpectedEnergyApprox}) 
[resp.  centered equation~(\ref{eqn:ivoExpectedEnergyApprox})].
Similarly, we let {\tt FTD} [resp. {\tt FTD}$^*$] denote the uncentered 
[resp. centered] version of our code from Algorithm 1 in 
Section ``Statistical Mechanics'' in Methods.  
In computing entropy for Rfam family RF00005,
both {\tt ViennaRNA} and {\tt ViennaRNA}$^*$ sometimes return entropy values
that are {\em larger} than the correct values computed by {\tt DP}, while
entropy values of {\tt FTD} [resp. {\tt FTD}$^*$] are always smaller
than [essentially always smaller] than those of {\tt DP}, 
as expected when using finite 
differences to approximate the derivative of the strictly decreasing,
concave-down function $\ln Z(T)$. Figure~\ref{fig:RNAfoldEntropy}
shows the distribution of entropy differences (DP-FTD, DP-FTD$^*$,
DP-ViennaRNA, DP-ViennaRNA$^*$) for 960 transfer RNAs from family
RF00005 from the Rfam 11.0 database \cite{Burge.nar13}.
%len	0.3725	0.3721	0.3688	1.0000	
%\footnote{Among those transfer RNAs whose
%{\tt DP} entropy is {\em less} than that computed by {\tt FTD}, 
%[resp. {\tt ViennaRNA}, {\tt ViennaRNA}$^*$] for RF00005, the average
%entropy difference is $-0.000656 \pm 0.000455$
%[resp. $-0.029937 \pm 0.028413$, $-0.018889 \pm 0.022221$].} 
Reasons for the behavior of
{\tt ViennaRNA} and {\tt ViennaRNA}$^*$ are presumably
due to numerical precision issues.
%RF00167
%DP-FTD Mean:0.037307	StDev:0.036789	Max:0.155959	Min:-0.001491
%DP-Vienna Mean:0.033700	StDev:0.054528	Max:0.169348	Min:-0.103386
%DP-ViennaStar Mean:0.033185	StDev:0.041779	Max:0.147751	Min:-0.054748
These differences are small, so when
plotted as a function of sequence length in a manner analogous to
Fig.~\ref{fig:runtimes} (not shown), average entropy values
computed by {\tt FTD}, {\tt FTD}$^*$, {\tt ViennaRNA}, and {\tt ViennaRNA}$^*$
for $\Delta T = 10^{-2}$ and $10^{-4}$ are visually indistinguishable. 
The left panel of
Fig.~\ref{fig:viennaRuntimes} shows that {\tt ViennaRNA} is somewhat faster
than {\tt FTD}, and for each method, the {\em uncentered} version is faster
than the {\em centered} version, which is clear since the former [resp.
latter] computes the partition function twice [resp. three times].
The right panel of 
Fig.~\ref{fig:viennaRuntimes} shows that the standard deviation of entropy
values for 100 random RNA is larger for {\tt ViennaRNA} than {\tt FTD}, and
the uncentered form of {\tt ViennaRNA} displays the largest standard deviation
when $\Delta T = 10^{-4}$ (for $\Delta T = 0.01$, all four finite derivative
methods are comparable).
These results are unsurprising due to numerical precision issues; e.g.
for the 98 nt purine riboswitch with EMBL accession code
AE005176.1/1159509-1159606, the algorithm {\tt DP} determines a value of
conformational entropy $9.975439$, whereas by using (centered)
{\tt ViennaRNA}$^*$ with version 2.1.8 of {\tt RNAfold} with $\Delta T = 10^{-2}$,
we obtain $9.93425742505$. For $\Delta T = 10^{-4}$, $10^{-5}$ and
$10^{-6}$, {\tt ViennaRNA}$^*$ computes entropies of 
9.59831636855, $6.94285165005 \cdot 10^{-8}$, $-5.9169597422 \cdot 10^{-7}$.
Such numerical instability issues are of much less concern to our method
{\tt FTD} and {\tt FTD}$^*$, as Fig.~\ref{fig:runtimes} demonstrates for
the uncentered method {\tt FTD} with $\Delta T = 10^{-7}$.

It follows that {\tt ViennaRNA} and {\tt ViennaRNA}$^*$ perform optimally with
$\Delta T = 10^{-2}$.
Note that when using {\tt RNAfold}, it is essential to use {\tt --betaScale};
indeed, if one attempts to compute the entropy using
equation~(\ref{eqn:entropyFTD}) where expected energy is computed from
equation~(\ref{eqn:tonoExpectedEnergyApprox})
[resp.  equation~(\ref{eqn:ivoExpectedEnergyApprox})]
by running {\tt RNAfold -p -T 37.01} and {\tt RNAfold -p -T 37} [resp.
{\tt RNAfold -p -T 37.01} and {\tt RNAfold -p -T 36.99}], then the resulting
entropy for the 98 nt purine riboswitch with EMBL accession code
AE005176.1/1159509-1159606 is the impossible, {\em negative} value of  
-208.13 [resp. -210.61].
The large negative entropy values in this case are not only due to the
lack of distinction between formal and table temperature, but as well
to the fact that Vienna RNA Package represents energies as integers
(multiples of 0.01 kcal/mol), so that loop energies jump at 
particular temperatures, as shown in the right panel of
Fig.~\ref{fig:RNAfoldEntropy}. These issues should not be construed as
shortcomings of the Vienna RNA Package, designed for great speed and
high performance, but rather as a use of the program outside its intended
parameters. As shown by Fig.~\ref{fig:viennaRuntimes}, the methods
{\tt ViennaRNA} and {\tt ViennaRNA}$^*$ can rapidly 
compute accurate approximations of the conformational entropy.

%Figure 5
%end Figure 5

%Figure 6
%end Figure 5

\subsection*{Correlation with hammerhead cleavage activity}
\label{section:hammerheadActivity}

In \cite{Shao.bb07}, Shao et al. considered a 2-state thermodynamic
model to describe the hybridization of hammerhead ribozymes to messenger
RNA with subsequent cleavage at the mRNA GUC-cleavage site. In that paper,
they define the total free energy
\begin{eqnarray}
\label{eqn:dingHammerhead1}
\Delta G_{\mbox{\small total}} &=&
\Delta G_{\mbox{\small hybrid}} -
\Delta G_{\mbox{\small switch}} -
\Delta G_{\mbox{\small disrupt}}
\end{eqnarray}
where each of these energies is defined on p. 10 of
\cite{Shao.bb07}, and obtained by averaging over 1000 low energy
structures sampled by {\tt Sfold}  \cite{Ding.nar04}.
The authors show a (negative) high correlation between 
$\Delta G_{\mbox{\small total}}$ and the cleavage activity of
13 hammerhead enzymes for GUC cleavage sites in 
ABCG2 messenger RNA (GenBank NM\_004827.2) of {\em H. sapiens}; i.e.
the lower the total change in free energy, the more active is the ribozyme.
(Shao et al. originally considered 15 hammerheads; however two 
outlier hammerheads were removed from consideration.)
Here, we show that the correlation with cleavage activity can be
improved slightly by taking secondary structure
conformational entropy into consideration.

\begin{figure}[!ht]
\centering
%\includegraphics[width=0.45\textwidth]{figure7a}
%\qquad
%\includegraphics[width=0.45\textwidth]{figure7b}
\includegraphics[width=\textwidth]{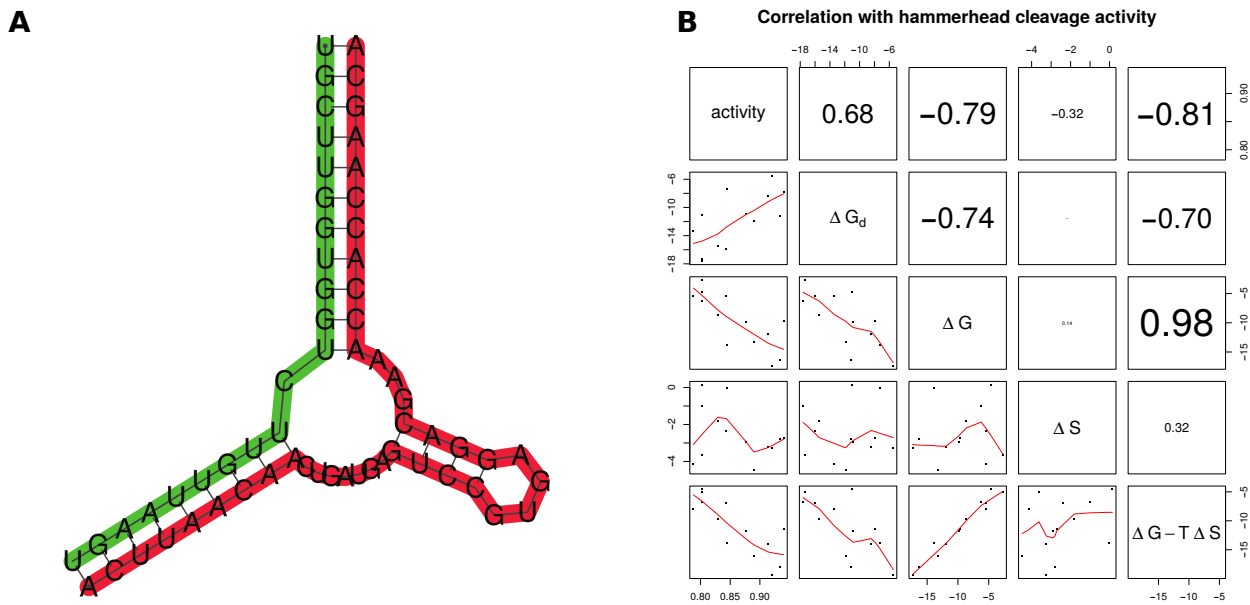}
\caption{{\em (A)}
%Secondary structure of a portion of the {\em H. sapiens} 
%ABC transporter ABCG2 messenger RNA (GenBank . NM\_004827.2) hybridized with
%a hammerhead ribozyme (data from the first line of Table 1 of
%\cite{Shao.bb07}). The 21 nt portion of mRNA is
%$5'$-UGCUUGGUGG UCUUGUUAAG U-$3'$
%and the 42 nt hammerhead rizozyme is
%$5'$-ACUUAACAAC UGAUGAGUCC GUGAGGACGA AACCACCAAG CA-$3'$.
%
Hybridization structure predicted by {\tt RNAcofold} 
\cite{stadler:Hybridization} of a 21 nt portion of
messenger RNA for {\em H. sapiens} 
ABC transporter ABCG2 messenger RNA (GenBank . NM\_004827.2) hybridized with
a hammerhead ribozyme (data from the first line of Table 1 of
\cite{Shao.bb07}). The 21 nt portion of mRNA is
$5'$-UGCUUGGUGG UCUUGUUAAG U-$3'$
and the 42 nt hammerhead rizozyme is
$5'$-ACUUAACAAC UGAUGAGUCC GUGAGGACGA AACCACCAAG CA-$3'$.
Messenger RNA is shown in green, while the hammerhead appears in
red. In data not shown, we determined the
secondary structure of the 21 nt mRNA portion, followed by a linker
region of 5 adenines, followed by the 42 nt hammerhead ribozyme, by
using {\tt RNAfold} \cite{Lorenz.amb11}. The base pairs in the
hybridization complex are identical to the base pairs in the 
chimeric single-stranded sequence (not shown) -- i.e. 
except for the unpaired adenines from the added linker region, the
structures are identical. This fact permits us to approximate the
structural entropy for the hybridization of two RNAs by using {\tt RNAentropy}
to compute the entropy of the concatenation of the sequences, separated
by a linker region.
{\em (B)}
Correlation between hammerhead cleavage activity, as assayed
by Shao et al. \cite{Shao.bb07}, with $\Delta G_d$ (change in free energy
due to disruption of mRNA, denoted $\Delta G_{\mbox{\small disrupt}}$ in text), 
$\Delta G$ (change in total free energy, denoted
$\Delta G_{\mbox{\small total}}$ in text),
both taken from \cite{Shao.bb07}, 
with $\Delta S$ (change in conformational entropy $k_B \cdot \Delta H$),
and $\Delta G$(total) - T $\Delta S$. Cleavage activity was measured by
Shao et al. for the cleavage of GUC sites in 
ABC transporter ABCG2 messenger RNA 
(GenBank NM\_004827.2). Values of $\Delta G_d$, $\Delta G$ were taken from 
Table~1 of \cite{Shao.bb07}, while the change in conformational entropy 
$\Delta S$ was computed by {\tt RNAentropy}.
Note modest increase in the correlation of cleavage activity with
$\Delta G$, when adding the free energy contribution $-T\Delta S$,
due to conformational entropy.
}
\label{fig:hammerheadActivity}
\end{figure}

To fix ideas, we consider the first GUC cleavage site considered by
Shao et al. The minimum free energy (MFE) hybridization complex, as predicted
by {\tt RNAcofold} from the Vienna RNA Package \cite{Lorenz.amb11} is
shown in Fig.~\ref{fig:hammerheadActivity}A.
The  MFE structure of the 21 nt portion of
mRNA, followed by a linker region of five adenines, followed by the
hammerhead ribozyme, as computed by {\tt RNAfold} from the Vienna RNA
Package yields the same structure (where the linker region appears in a
hairpin). It follows that
to a first approximation, MFE hybridization structures
can be predicted from MFE structure predictions of a chimeric sequence
that includes a linker region. (Before the introduction of hybridization
MFE software \cite{Dimitrov:2004lr,Lorenz.amb11}, this approach was used
to predict hybridization structures.)

In this case, enzyme activity is $0.843$,
$\Delta G_{\mbox{\small total}} = -5.423$ kcal/mol,
structural entropy of the hammerhead is 2.830,
structural entropy of the 21 nt portion of mRNA is 2.146, and
structural entropy of the 
21 nt portion of mRNA portion with linker and hammerhead is 2.328.
Assuming that the entropy of a rigid structure is zero, the change
in structural entropy $\Delta H$(hammerhead) is $0-2.830 = -2.830$,
and similarly $\Delta H$(21 nt mRNA + linker) is $-2.146$, 
$\Delta H$(21 nt mRNA+linker+hammerhead) is $-2.328$. The net change
in structural entropy 
$\Delta H$ is 
$\Delta H$(21 nt mRNA+linker+hammerhead) minus
$\Delta H$(21 nt mRNA + linker) minus
$\Delta H$(hammerhead), so $\Delta H = -2.328 -(-2.146 - 2.830) = 2.648$.
The net change in conformational entropy $\Delta S = k_B \cdot \Delta H$
is then $0.00526$, hence the free energy contribution
$-T \Delta S = -RT \Delta H = -1.632.$ The correlation
between $\Delta G_{\mbox{\small total}}$ and $-T \Delta S$ is the value
of $0.108$, while the correlation value of $-0.788$ 
between hammerhead activity and $\Delta G_{\mbox{\small total}}$  is
increase in absolute value to $-0.806$ (p-value $0.000878$)
when also taking into account $-T \Delta S$. 
See Fig.~\ref{fig:hammerheadActivity} for a scatter plot and
correlations between enzyme activity and $\Delta G$
[resp. $\Delta G - T \Delta S$], which correspond to the
total free energy change 
without [resp. with] a contribution from conformational entropy.
%See Table~\ref{table:hammerheadActivity} for full details.

%Figure 7

Fig.~\ref{fig:hammerheadActivity}A  depicts the
minimum free energy {\em hybridization} structure of a 21 nt portion
of the ABC transporter ABCG2 messenger RNA from {\em H. sapiens}
(GenBank NM\_004827.2), 
hybridized with a hammerhead ribozyme (data from the first line of Table 1 of
\cite{Shao.bb07}). The MFE hybridization structure was computed by
Vienna RNA Package {\tt RNAcofold} \cite{Lorenz.amb11}. We obtain the
same structure by applying {\tt RNAfold} to the chimeric sequence obtained
by concatenating the 21 nt portion of mRNA, given by
$5'$-UGCUUGGUGG UCUUGUUAAG U-$3'$, with a 5 nt linker region consisting of
adenines, with the 42 nt hammerhead rizozyme, given by
$5'$-ACUUAACAAC UGAUGAGUCC GUGAGGACGA AACCACCAAG CA-$3'$ (data not shown).
By such concatenations with a separating 5 nt linker region, we can
compute the structural entropy of hybridizations of the 21 nt mRNA with
the hammerhead ribozyme. (In future work, we may
extend {\tt RNAentropy} to compute the entropy of
hybridization complexes without using such linker regions.)

\subsection*{Structural entropy of HIV-1 genomic regions}
\label{section:hiv}

\begin{figure}[tbph]
\centering
\includegraphics[width=1.00\textwidth]{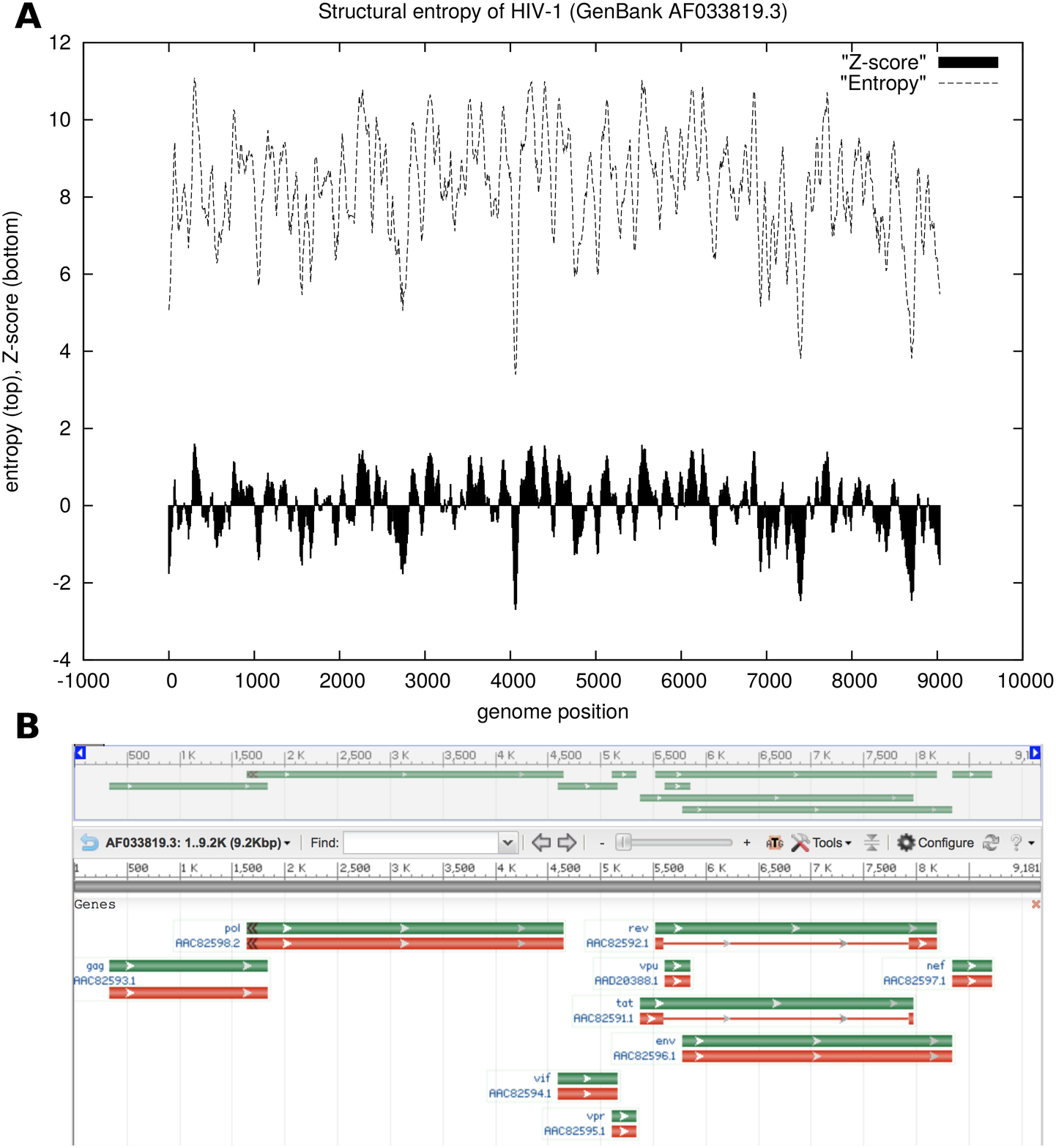}
\caption{Structural entropy plot for the HIV-1 genome
(GenBank AF033819.3). Using {\tt RNAentropy}, the structural
entropy was computed for each 100 nt portion of the HIV-1 genome, 
by increments of 10 nt; i.e. for 100 nt windows starting at genome
position 1, 11, 21, etc. To smooth the curve, moving averages were
computed over five successive windows.
{\em (A)} Dotted-line displays moving average values of structural
entropy; solid curve displays entropy Z-scores, defined by 
$\frac{x-\mu}{\sigma}$, where $x$ represents the (moving window average)
entropy at a genomic position, and $\mu$ [resp. $\sigma$] represents
the mean [resp. standard deviation] of the entropy for 100 nt windows.
Some of the lowest entropy Z-scores are -2.69 at position 4060, -2.46 
at position 8700, -1.95 at position 4040.
{\em (B)} NCBI graphics display of the HIV-1 genome, for comparison
purposes. Low entropy (negative Z-score) regions do not appear to 
correspond with the start/stop location for annotated genes.
In data not shown, we also computed {\em positional entropy} values
\cite{Huynen.jmb97} for the same windows, and determined a 
Pearson correlation of 0.7025 [resp. Spearman correlation of 0.6829]
between (moving window average) values of entropy and positional entropy.
}
\label{fig:hiv1entropy}
\end{figure}

Fig.~\ref{fig:hiv1entropy}A depicts the structural entropy, computed
as a moving average of 100 nt portions of the HIV-1 complete genome
(GenBank AF033819.3). Using {\tt RNAentropy}, the structural
entropy was computed for each 100 nt portion of the HIV-1 genome,
by increments of 10 nt; i.e. entropy was computed at genomic positions
1, 11, 21, etc. for 100 nt windows. To smooth the data, moving averages were
computed over five successive windows. The figure displays the moving
average entropy values, as a function of genome position (top dotted curve),
entropy Z-scores, defined by
$\frac{x-\mu}{\sigma}$, where $x$ is the (moving window average)
entropy at a genomic position, and $\mu$ [resp. $\sigma$] is
the mean [resp. standard deviation] of the entropy for all computed
100 nt windows. Fig.~\ref{fig:hiv1entropy}B is a portion of the
NCBI graphics format presentation of GenBank file AF033819.3.
Regions of low Z-score are position 4060 (Z-score of -2.69), 
position 8700 (Z-score of -2.46) and position 4040 (Z-score of -1.95). 
Since positions do not appear to 
correspond to the start/stop position of annotated genes,
we ran {\tt cmscan} from {\tt Infernal 1.1} software \cite{Nawrocki.b13} on
the HIV-1 genome (GenBank AF033819.3). We obtained 
11 predicted noncoding elements as listed in Table~\ref{table:hiv1elements},
including the trans-activation response (TAR) element. Many of the 
predicted noncoding RNAs are much shorter than the 100 nt window used in
the {\tt RNAentropy} genome-scanning approach just described -- it follows
that low entropy Z-scores cannot be expected for such elements.
Nevertheless, certain elements have quite low 
entropy Z-scores, such as the $5'$-UTR and TAR element, both of which are 
known to be involved in the packaging of two copies of the HIV-1 genome 
in the viral capsid \cite{Lu.jmb11}. 

%Figure 8
%end Figure 8

%Table 6
\begin{table}[!ht]
\caption{Computationally annotated RNA noncoding elements
from the HIV-1 genome with corresponding entropy Z-scores.
Running {\tt cmscan} from {\tt Infernal 1.1} \cite{Nawrocki.b13} on
the HIV-1 genome (GenBank AF033819.3),
we obtain 11 noncoding elements as listed in the table, along
with the nucleotide beginning and ending positions, length of
noncoding element, E-score, and entropy Z-score. Entropy Z-scores were
computed using {\tt RNAentropy} as explained in the text. Many of
the annotated noncoding elements are much shorter than 100 nt, the length
of the window size used; however, sporadic checking of entropy Z-scores
computed for a moving window of size 50 does not seem to radically change the
entropy Z-scores. Nevertheless, certain elements have low entropies
and corresponding entropy Z-scores, such as the $5'$-UTR and 
TAR (trans-activation response) element, both of which are known to be
involved in the packaging of the HIV-1 genome in the viral
capsid \cite{Lu.jmb11}.
}
\begin{tabular}{|l|rrrrr|}
\hline
Name & Start & Stop & Len & E-score  & entropy Z-score \\
\hline
RRE  & 7265  & 7601 & 66  &  7.6e-125 & -1.389 \\
HIV PBL & 125 & 223 & 99 & 1.6e-30 & -0.589 \\
HIV POL-1 SL & 2012  & 2124 & 113 & 3.1e-29  & +0.066 \\
HIV GSL3  & 400  &  483 & 84 & 1.2e-23 & -0.299 \\
mir-TAR   & 9085 & 9145 & 61 & 7e-21  & -1.528\\
mir-TAR   &1     &60  & 60 & 1.1e-18 & -1.759\\
HIV FE   & 1631  & 1682 & 52 & 3.6e-11  &  -0.506\\
HIV-1 DIS & 240  & 279 & 40 & 3.7e-11 &  -0.205\\
HIV-1 SL3  &309 & 331 & 23 & 7.1e-09 & +0.907\\
HIV-1 SL4  & 337  & 356 & 20 & 1.9e-05 & +0.907\\
HIV-1 SD & 282 & 300 & 19 & 3.7e-05  &  -0.529 \\
\hline
\end{tabular}
\label{table:hiv1elements}
\end{table}
%end Table 6

\section*{Discussion}
\label{section:discussion}

In this paper, we have introduced two cubic time algorithms, both
implemented in the publicly available program {\tt RNAentropy}, to
compute the RNA {\em thermodynamic structural entropy},
$H = - \sum_s p(s) \ln p(s)$, where $p(s) = \exp(-E(s)/RT)/Z$ is the
Boltzmann probability of secondary structure $s$, 
and the sum is taken over all structures of a given RNA
sequence $\aseq = a_1,\ldots,a_n$. This answers a question raised
by M. Zuker (personal communication, 2009).  Taking a benchmarking
set that consists of the first RNA from each of the 2450 families from
database Rfam 11.0 \cite{Burge.nar13}, we determined the 
correlation of thermodynamic structural entropy with a variety of
other measures used in the computational design and experimental
validation of synthetic RNA \cite{Zadeh.jcc11,Dotu.nar15}. 

In  \cite{Manzourolajdad.jtb13},  Manzourolajdad et al. described an
algorithm to compute RNA structural entropy
$H = -\sum_s p(s) \ln p(s)$, where $p(s)$ is the probability of the 
(unique) leftmost derivation of the sequence-structure pair $\aseq,s$,
{\em conditioned} on the probability of deriving the sequence $\aseq$.
Using random RNA, the 960 seed alignment sequences from Rfam family RF00005, 
and a collection of 2450 sequences obtained by selecting the
first RNA from the seed alignment of each family from
the Rfam 11.0 database \cite{Burge.nar13}, 
we show the following:
(1) the thermodynamic structural entropy algorithms DP, FTD compute the same 
structural entropy values with the same efficiency, although as sequence
length increases, FTD runs somewhat faster and returns slightly smaller values 
than does DP.
(2) DP and FTD appear to be an order of magnitude 
faster than the SCFG method of \cite{Manzourolajdad.jtb13}, which latter
requires two minutes for RNA sequences of length 500 that require only a
few seconds for DP and FTD.
(3) Derivational entropy values computed by the
method of \cite{Manzourolajdad.jtb13} are much larger than thermodynamic
structural entropy values of DP and FTD,
ranging from about 4-8 times larger,depending on the SCFG chosen.
(4) The length-normalized
correlation between thermodynamic structural entropy values and
derivational entropy values is poor to moderately weak.

Why are SCFG structural entropy values much larger than thermodynamic
structural entropy values knowing that all entropies are computed with the natural
logarithm?   Indeed,
by numerical fitting of DP and SCFG entropy values for random RNA depicted in 
Fig.~\ref{fig:runtimes}B, we determine that SCFG(G6) entropy values
are 3.56 times larger than DP,
while G4 and G5 entropy values are 6.71 resp.  6.85
times larger than DP entropy values.
From results and discussion in
\cite{Anderson.bb13,Sukosd.bb13},  one might 
speculate that derivational entropy of a given RNA sequence $\aseq$
might be smaller if the SCFG correctly captured the `essence' of 
particular training set of RNAs, and that $\aseq$ resembles the RNAs
of the training set. However this cannot be correct, since
Fig.~\ref{fig:lengthNormEntropy}B presents
derivational entropy values for 960 transfer RNAs from family
RF00005 from the Rfam 11.0 database \cite{Burge.nar13}, where grammars
G4, G5 and G6 were {\em trained} on the dataset `Rfam5'. At present,
there is no clear  answer to the question of why derivational entropy
values are so much larger than thermodynamic structural entropy values.
At the very least, the difference in entropy values indicates that
secondary structures have very different probabilities, depending on
the algorithm used.

We now discuss the relation with work of Miklos et al. \cite{Miklos.bmb05},
who described a dynamic programming
algorithm to compute the expected energy of an RNA sequence.
In personal communications, the main authors,
I. Miklos and I.M. Meyer, have both reported that their original dynamic
programming code appears to be lost. Moreover, only the general idea 
\[
Q_{i,j} = Q_{i,j-1} + \sum_{k=i}^{j-4} bp(k,j) \left[
Q_{i,k-1} Z_{k,j} +
Z_{i,k-1} Q_{k,j}  \right]
\] 
of their algorithm is described in \cite{Miklos.bmb05}, corresponding
approximately, but not exactly, with 
equation~(\ref{eqn:nussTurnerExpNhbors}) in Section~\ref{section:entropyDP}
\[ 
Q_{i,j} = Q_{i,j-1} + \sum_{k=i}^{j-4} bp(k,j) \left[
Q_{i,k-1} ZB_{k,j} +
Z_{i,k-1} QB_{k,j}  \right].
\]
In particular, none of the explicit details of Section~\ref{section:turner}
concerning the 
recursions for treating hairpins, bulges, internal loops, and multiloops
appear in \cite{Miklos.bmb05}.  For these reasons, we developed our own
recursions and implemented our own DP algorithm to compute expected energy.
As shown in Fig.~\ref{fig:runtimes}, it takes only a few seconds to 
compute the entropy of an RNA sequence of length 500 nt on a
Core2Duo PC (2.8 GHz; a 2 Gbyte memory; CentOS 5.5). In contrast,
Miklos et al. \cite{Miklos.bmb05} state that their code took about
10 minutes to compute the entropy and {\em variance} for an RNA sequence
of length 120 nt, using a Pentium4 2.0 GHz computer. As the
presumably slower program of Miklos et al. is no longer available, 
the public availability of our program {\tt RNAentropy} may be of benefit
to other researchers.

In \cite{Salari.nar13} Salari et al. describe a dynamic programming algorithm
to compute the {\em relative entropy}, or Kullbach-Liebler distance,
$P || Q$, where $P$ is the 
Boltzmann probability distribution for all secondary structures of a given
RNA sequence, and $Q$ is the 
Boltzmann probability distribution for all secondary structures of single
point mutant of that sequence (an energy assumption is made to avoid zeros in
the denominator when computing relative entropy). The recursions given in
\cite{Salari.nar13} are similar to but distinct from those given in the
current paper, and to our knowledge, the software of Salari et al.,
which would need modification to compute entropy,  is not available.

There are three future additions that may make our code, {\tt RNAentropy},
more useful. First, it is possible to extend the code in order to
compute expected energy and the structural entropy of a hybridization
complex. Second, it is possible to incorporate {\em hard constraints},
where all structures are required to have certain positions base-paired
together, or certain positions to be unpaired. Such hard constraints were
first introduced in \cite{mathews:RNAstructure}. Third, it is possible to
incorporate {\em soft constraints}, where Boltzmann weights penalize
positions that deviate from chemical footprinting data,
such as in-line probing or
selective 2'-hydroxyl acylation analyzed by primer extension (SHAPE).
For details on soft constraints, see Zarringhalam et al.
\cite{Zarringhalam.po12}, as well as the web server
\url{http://bioinformatics.bc.edu/clotelab/RNAsc}. Although entropy can
be computed using a simple script that calls
{\tt RNAfold -p --betaScale}, both hard and soft constraints are handled
quite differently by the Vienna RNA Package, so the suggested enhancements of
{\tt RNAentropy} may prove useful.

Our program, {\tt RNAentropy}, has two versions, depending on whether the
user wishes to use the Turner 1999  parameters, or the newer Turner 2004
parameters \cite{Turner.nar10} (in both cases, energy parameters do not
include dangle or coaxial stacking, and were obtained from the Vienna RNA
Package \cite{Lorenz.amb11}). Additionally, {\tt RNAentropy} implements
the method described in Section~\ref{section:statisticalPhysics}, which
computes expected energy 
$\langle E \rangle = RT^2 \cdot \frac{\partial}{\partial T} \ln Z(T)$,
by {\em uncoupling} formal and table temperatures. For convenience, we
also make available a script to compute entropy by calling 
{\tt RNAfold --betaScale}.  For relatively short RNAs, the uncentered
formal temperature derivative method is fast and accurate,
as implemented in methods {\tt FTD} and {\tt ViennaRNA}, while the
centered versions {\tt FTD}$^*$ and {\tt ViennaRNA}$^*$ are somewhat slower.
Since {\tt Vienna RNA Package} has been under constant development,
refinement and extension for approximately 30 years, the software enjoys
an efficiency and speed that is remarkable (see 
Fig.~\ref{fig:viennaRuntimes}A).
When using methods {\tt ViennaRNA} and {\tt ViennaRNA}$^*$, it is recommended
to use $\Delta T = 10^{-2}$ since smaller values lead to increasingly
incorrect values. In contrast, {\tt FTD} and {\tt FTD}$^*$ may be used
with $\Delta T$ as small as $10^{-7}$, although when benchmarking against
random RNA of length 20-500, the data (not shown)
suggest that differences between {\tt DP} and {\tt FTD} 
[resp. {\tt FTD}$^*$] are minimized for $\Delta T = 10^{-2}$ [resp.
$\Delta T = 10^{-9}$] (nevertheless, the choice of $\Delta T$ makes little
difference for {\tt FTD} and {\tt FTD}$^*$).
For larger sequences, real entropy values, as computed by {\tt DP} exceed 
the approximate methods by a larger margin, hence we recommend that {\tt DP} 
should be used.
Our code is available at 
\url{http://bioinformatics.bc.edu/clotelab/RNAentropy}.

\section*{Acknowledgments}
We would like to thank I.L. Hofacker for kindly pointing out the
flag {\tt --betaScale} in {\tt RNAfold}, and for suggesting to consider
a centered finite difference when approximating the derivative of
$\ln Z$.
We would also like to thank both I. Miklos and I.M. Meyer for correspondence
concerning the availability of their code from \cite{Miklos.bmb05}, 
Y. Ding for generously providing the hammerhead and mRNA data from
\cite{Shao.bb07}, and E. Rivas for a discussion about SCFGs.
%This research was funded by the National Science Foundation grant DBI-1262439.
%Any opinions, findings, and conclusions or recommendations expressed in 
%this material are those of the authors and do not necessarily reflect 
%the views of the National Science Foundation.

\bibliographystyle{plain}
%\bibliography{biblio}

\end{document}